\documentclass[preprint,12pt]{elsarticle}

\usepackage{amssymb}
\usepackage{amsmath} % for align
\usepackage{listings}% for code listings
\usepackage{xcolor}  % to set colors in code listings
\usepackage{siunitx} % for proper units
\usepackage{upgreek} % for upbeta
\usepackage{mathtools} % for abs
\usepackage{booktabs} % for nicer tables
\usepackage{url} % for linebreaks bib item urls

\usepackage[pagewise]{lineno}

\DeclareSIUnit\statcoulomb{statC}
\DeclareSIUnit\statvolt{statV}
\DeclareSIUnit\erg{erg}

\newcommand\changed[1]{#1}
\newcommand\cchanged[1]{#1}

\colorlet{punct}{red!60!black}
\definecolor{background}{HTML}{EEEEEE}
\definecolor{delim}{RGB}{20,105,176}
\colorlet{numb}{magenta!60!black}

\lstdefinelanguage{json}{
    basicstyle=\footnotesize\ttfamily,
    numbers=left,
    numberstyle=\scriptsize,
    stepnumber=1,
    numbersep=8pt,
    showstringspaces=false,
    breaklines=true,
    frame=lines,
    backgroundcolor=\color{background},
    literate=
     *{0}{{{\color{numb}0}}}{1}
      {1}{{{\color{numb}1}}}{1}
      {2}{{{\color{numb}2}}}{1}
      {3}{{{\color{numb}3}}}{1}
      {4}{{{\color{numb}4}}}{1}
      {5}{{{\color{numb}5}}}{1}
      {6}{{{\color{numb}6}}}{1}
      {7}{{{\color{numb}7}}}{1}
      {8}{{{\color{numb}8}}}{1}
      {9}{{{\color{numb}9}}}{1}
      {:}{{{\color{punct}{:}}}}{1}
      {,}{{{\color{punct}{,}}}}{1}
      {\{}{{{\color{delim}{\{}}}}{1}
      {\}}{{{\color{delim}{\}}}}}{1}
      {[}{{{\color{delim}{[}}}}{1}
      {]}{{{\color{delim}{]}}}}{1},
}

\newcommand{\fullstop}{\text{\,.}} % in equations
\newcommand{\commaeq}{\text{\,,}} % in equations
\DeclarePairedDelimiter\abs{\lvert}{\rvert} % for \abs

\newcounter{bla}

\journal{Computer Physics Communications}

\begin{document}
%\linenumbers  
\begin{frontmatter}

  \title{KARL - A Monte Carlo model for atomic and molecular processes in the tritium atmosphere of the KATRIN experiment}

	\author[a]{Christian Sendlinger}
  \author[b]{Jonas Kellerer}
  \author[a]{Felix Spanier \corref{author}}

  \cortext[author] {Corresponding author.\\\textit{E-mail address:} felix@fspanier.de}
  \address[a]{Institut f\"ur Theoretische Astrophysik, Universit\"at Heidelberg, Albert-Ueberle-Str. 2 und Philosophenweg 12,
    69120 Heidelberg, Germany}
  \address[b]{Institut f\"ur Astroteilchenphysik, KIT, Hermann-von-Helmholtz-Platz 1,
    76344 Eggenstein-Leopoldshafen, Germany}
  
  \begin{abstract}

\changed{A new parallelized simulation code is presented, which uses a Monte Carlo method to determine particle spectra in the KATRIN source. Reaction chains are generated from the decay of tritium within the source. The code includes all relevant processes: elastic scattering, ionization, excitation (electric, vibrational, rotational), recombination and various clustering processes. The main emphasis of the code is the calculation of particle spectra and particle densities and currents at specific points within the source. It features a new technique to determine these quantities. It also calculates target fields for the interaction of particles with each other as it is needed for recombination processes.\\
The code has been designed for the KATRIN experiment but is easily adaptable for other tritium based experiments like Project 8. Geometry and background tritium gas flow can be given as user input.\\
The code is parallelized using MPI and writes output using HDF5. Input to the simulation is read from a JSON description.% that provides general information on the simulation (resolution of the simulation domain).% and the various optional features of the code, namely user defined species with their own terminator, interaction types, movers and initialization.
 }


{\bf PROGRAM SUMMARY}

\begin{small}
  \noindent
  {\em Manuscript Title:} KARL - A Monte Carlo model for atomic and molecular processes in the tritium atmosphere of the KATRIN experiment \\
  {\em Authors:} Jonas Kellerer, Christian Sendlinger, Felix Spanier \\ % Reiling hier drinnen lassen, auch wenn code nicht mehr von ihm enthalten ist?
  {\em Program Title:} KARL - \textbf{KA}trin WGTS elect\textbf{R}on and ion spectrum Monte Car\textbf{L}o                                \\
  {\em Journal Reference:}                                      \\ %Leave blank, supplied by Elsevier.
  {\em Catalogue identifier:}                                   \\ %Leave blank, supplied by Elsevier.
  {\em Licensing provisions:} GNU Public License                \\ %muss so sein wegen blitz
  {\em Programming language:} C++                               \\
  {\em Computer:} any workstation or cluster that has a modern C++ compiler (e.g. g++ 8.2 or later) an MPI implementation and the required external libraries \\
  {\em Operating system:} Linux / Unix                          \\
  {\em RAM:} depending on problem size and number of resolved species between 100 MB and 2 GB;  some scratch space \\
  {\em Number of processors used:} depending on problem size between one and a few hundred processors 
  {\em Classification:} 19.1 Atomic and Molecular Processes              \\ %Classify using CPC Program Library Subject Index, see ( % http://cpc.cs.qub.ac.uk/subjectIndex/SUBJECT_index.html)
  {\em External routines/libraries:} C++ compiler (tested with g++ 8.2 and 9.4.0), MPI 1.1 (tested with OpenMPI 3.1), HDF5 with support for parallel I/O (tested with version 1.10.0), Blitz++ (tested with version 1.0.2), Jansson (tested with version 2.12 and 2.13)\\
  {\em Nature of problem:} In the KATRIN experiment (and other experiments alike that feature large vessels filled with tritium) electrons are created from beta decay. These electrons interact with the ambient gas to produce secondary electrons through ionization. Subsequent processes include excitation, secondary ionization and  collisions. The resulting electron and ion differential energy spectrum at various positions is relevant for further plasma analysis, and the current of charged particles to the ends of the experiments is an observable. \\
  {\em Solution method:} Semi-classical Monte Carlo \\
  {\em Restrictions:}  The geometry of the experiment is currently limited to the KATRIN experiment, but this may easily be changed.\\
  {\em Unusual features:} The configuration is stored in JSON files. \\
  {\em Running time:} minutes to hours; depending on number of simulated decays \\

\end{small}
\end{abstract}
\end{frontmatter}
\section{Introduction}
\label{sec:intro}

One of the open questions in modern particle physics is the mass of the neutrino. Although the observation of neutrino oscillations clearly shows a non-zero neutrino mass, a specific mass could not be determined. One of the current experiments for the determination of the neutrino mass is KATRIN \citep{2001hep.ex....9033K}. The general measuring principle of KATRIN is the observation of the electron spectrum of tritium beta decay near the endpoint. The spectrum near the endpoint depends on the actual neutrino mass. The KATRIN experiment is designed to measure the electron anti-neutrino mass to unprecedented sensitivity down to \SI{0.2}{eV} \citep{KATRIN2005}. 

The experiment can be divided into two parts: the source section and the detector section. In the source section, tritium gas is circulated which provides a high luminosity stream of beta electrons. These electrons are magnetically guided to the detector section. There they are separated by their kinetic energy: one small stream being recorded by the detector, the other larger one reflected back to the source. Both the initial and reflected stream of electrons interact with the gas in the source. \changed{These interactions include among others elastic scattering, ionization and excitation processes.}

The atomic and molecular processes produce secondary particles which may subsequently yield even more particles, which are either eliminated through recombination or similar processes or absorbed by the walls of the experiment. The charged particle density is high enough, that a plasma forms in the source \citep{Kuckert.2016}. \cchanged{This plasma influences the shape of the energy spectrum through its self generated inhomogeneous electric potential. This also includes the electrons that are above the detection threshold of \SI{18.6}{\kilo\eV}.} A precise knowledge of the electron and ion energy spectrum and density is necessary for any plasma analysis. 

Two different methods of descriptions seem obvious: a description through the temporal evolution of phasespace elements and a description through the Monte Carlo method by tracking of single particles \citep{MOBUS2001519}. \changed{The first method impresses with a high resolution in space and velocity but is very computational intensive, especially with many different interaction types and species. A comparison with Monte Carlo type simulations is complicated: Since this would not be a statistical method the noise is negligble, but the requirements to save the phase-space distribution (specifically for high velocities) are extremely high. Thus, it was rejected for the simulation of the problem at hand. However, in a classical Monte Carlo scheme each of the simulated events is independent of the others. This is not possible if direct interactions (e.g. recombination) between electrons and ions are of interest.} This problem is addressed in the KARL code, which is presented in the this paper.

\changed{The simulation approach realized through the KARL code works under the assumption that all atomic and molecular processes are fast compared to the plasma processes. Using a Monte Carlo like approach the effects of different processes are taken into account and energy spectra of different species are produced. These spectra will be utilized to calculate plasma potentials elsewhere. Resulting plasma potentials can then be fed back into the KARL code to obtain refined particle distributions.} 

\changed{The experimental setup of the KATRIN experiment is used as the default setup in the code. Other experiments using tritium decay may easily be implemented as the geometry and injection characteristics can be modified. This applies particularly to the  Project 8 experiment \citep{2022arXiv220307349P}. Within the KARL framework the majority of all tritium reactions are implemented, but some reactions that occur in vastly different parameter regimes may not be implemented yet.}\newline
\cchanged{Within the Project 8 experiment atomic tritium is used instead of molecular one. In this case (or a similar cases in which new particle species are needed) some adjustments to the code have to be made. First and foremost the new particle specie has to be included. This, however, only requires small changes within the code\footnote{It is necessary to define a new \textit{ImplementedSpecie} in the `species.h' file, 	define an alias and its charge in `species\_generator.h' and its mass in `mass\_generator.h'. }. Secondly it is necessary to include the corresponding cross sections with the new particle specie. This is either possible by providing the cross section as an external input, which does not require a changes within the code, or by implementing the analytic expression\footnote{The implementation should be done in `cross\_sections.cpp'. In order to be able to make use of the newly implemented function it is necessary to adjust the `interaction\_generator.cpp' file. Here the already existing cross sections serve as a guide for the new one.}. Apart from the change in cross section the numerical procedures for the interactions do not change. The possible interactions with atomic tritium includes almost all processes that are possible with molecular tritium (only rotational excitation is no longer possible).}

\section{Description of the method}
\label{sec:method}

\changed{In the KARL code, particles are tracked from interaction to interaction. The distance between the interactions is determined from the mean free path (MFP) to sample a Poisson process. The performed interaction is then drawn from a user-specified list of possible interactions. During the simulation new particles are created either through beta decay or through interaction between particles. All particles are put into a list, from where they are processed until they terminate. In case the list is empty new decay particles are created and inserted. The energy distribution of electrons created by the decay process is relatively simple to describe, since it follows the Fermi statistics of beta decay \citep{Fermi:1934aa}.}
%\changed{The general setup of the code is such that in the first place incident particles from tritium beta decay are tracked. In the simulation decay events are successively processed. Meaning that each decay particle is followed until its destruction, during which it might produce new, secondary particles. All primary and secondary particles are processed until termination before new decay particles are added.
%The initial energy spectrum for electrons created by the tritium beta decay is comparably simple to describe, as it follows the Fermi statistics of beta decay \citep{Fermi:1934aa}. 
%In the KARL code, single particles are tracked through the source from one interaction to another. The moved distance between interactions is determined from the mean free path. The performed interactions are drawn from a list of available, user-specified interactions.
%Through these interactions decay particles form a reaction tree with subsequently generated secondary particles. %(or with pre-existing particles that change properties via the interaction). 
%In order to follow all interactions of this tree each secondary particle forms a new ``branch''. 
%Newly generated particles are put into a list that is processed after the incident particle reaches the end of its lifetime.}

\changed{Apart from the decay process new particles are created mainly through electron impact ionization.}
\changed{Ionization produces two indistinguishable electrons with energies between \SI{0}{eV} and the impact energy subtracted by the ionization energy.}  

\changed{For electrons below \SI{100}{eV} elastic scattering is the dominant process, while for electrons above that energy  ionization becomes dominant, see Figure~\ref{fig:et2_interactions}. Depending on the neutral gas density this can lead to a large amount of secondary electrons being created.}

%For high energy electrons these interactions include mostly elastic scattering, but also ionization. Hence, many more electrons, and therefore also ions, are created in the source than from beta decay alone. 

\changed{The incident and secondary particles may produce a feedback to other particles - primarily through recombination and clustering processes:} Electrons can recombine with the existing ions in the source. This process is not only dependent on the electron spectrum, but the ion spectrum as well. There are many different types of ions, which are created by clustering processes with the neutral gas. The neutral gas density, which is the dominant interaction partner for both electrons and ions, is not constant in the source, but ranges over more than three orders of magnitude over \SI{16}{m} \citep{Kuckert.2018}. Additionally, it has a position dependent mean drift velocity, also covering two orders of magnitude \citep{Kuckert.2018}. 

The KARL code is customized to the special conditions in the KATRIN source. Hence, special care is taken to ensure correct movement of particles through the large density gradient. Additionally, all relevant interactions in a tritium gas atmosphere are included for both electrons and ions.

\subsection{General diagram}
\label{ssec:general_diagram}

\begin{figure}[tbp]
  \centering
  \includegraphics[height=0.8\textheight]{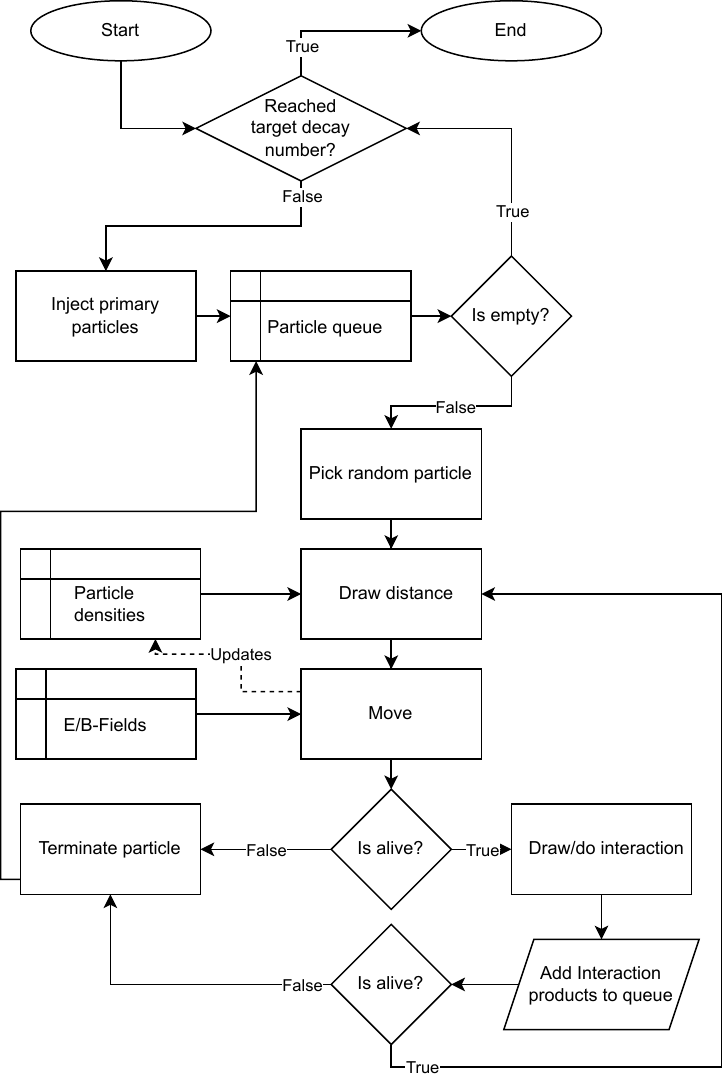}
  \caption{After initialization the simulation follows this flow diagram on each CPU core. See section~\ref{ssec:parallelization} for more information on the parallelization strategy.}
  \label{fig:simulation_flow}
\end{figure}

\changed{The main simulation cycle of the code consists of three subcycles, as shown in Figure~\ref{fig:simulation_flow}.}
%After initialization (see section~\ref{ssec:initialization}), the simulation starts its iterative cycle, which contains three subcycles, see Figure~\ref{fig:simulation_flow}.
The first addresses how many primary particles/events are processed, which is predefined by the user. It adds primary particles to the particle list, if it is empty. In the second subcycle a particle is selected at random from the particle list. This particle is processed until its termination and a new particle is selected from the list. 

Within the third subcycle, the \changed{particle movement and any interactions are performed}.
%The particle tracking is divided into two alternating steps. First, the particle is moved according to the electromagnetic fields in the source and the mean free path of the particle. Secondly, an interaction is performed.

For each movement step, the mean free path $\lambda_\text{mfp}$ of the corresponding particle is calculated through \citep{hussein2010radiation}
\begin{equation}
\label{eq:mean_free_path}
  \lambda_\text{mfp} = \frac{1}{\sum_i \sigma_i(E) n_i} \commaeq
\end{equation}
where $E$ is the kinetic energy of the particle in the \changed{center of mass} frame, $\sigma_i$ is the cross section of a specific interaction, and $n_i$ the particle number density of the target specie of the interaction. In the lab frame, the neutral gas has a non-zero drift velocity. The cross sections are evaluated from predefined functions. \changed{The particle number density is either calculated during the simulation for particle species that are tracked or through user input at the beginning of the simulation for particle species that act as a background population}.
%The particle number density of the interaction partner is calculated either from a predefined function for background species, or from the density field for tracked species, see section~\ref{ssec:denstiy_fields} and \ref{sec:implementation} for more information. 
The actual moved distance $d$ is a statistical quantity, which \changed{is} derived from the MFP through
\changed{
\begin{equation}
  d = -\lambda_\text{mfp} \cdot \log \eta \commaeq
  \label{eq:distance_from_mfp}
\end{equation}}
where $\eta \in (0, 1]$ is an uniform random number. 

\changed{This results in an exponential distribution for the distance $d$. This exponential distribution follows from the assumed Poisson distribution of the number of interactions within a certain path length.}
%
%This is due to an assumed Poisson random process, that results in an exponential probability distribution. 

The particle is then moved by this distance $d$ \changed{taking} the electromagnetic fields inside the source \changed{into consideration}. %, see section~\ref{ssec:movement}. 
During this step the density fields of the corresponding specie are updated.

After the movement step it is evaluated if the particle needs to be terminated, which is mainly a check if the particle has left the simulation domain. %, or has hit a wall, see~\ref{ssec:termination}. 
If the particle is still alive, a random interaction is selected from the list of available interactions, weighted by its probability to occur. The probability scales with the cross section of that interaction and the interaction partner particle density. The corresponding interaction is \changed{then} performed and the interaction products, \changed{not including the incident particle}, are added to the particle list. If the particle does no longer exist after the interaction (through recombination or clustering) it is terminated, otherwise the next movement step is performed.

\changed{The details of the implementation of the described components are described later in section~\ref{sec:implementation}.}

\subsection{Density fields}
\label{ssec:denstiy_fields}

In a general Monte Carlo simulation, each event is assumed to be independent of the other. However, different particles types can in general interact with each other, for example electrons and ions through recombination. Therefore, each event in our simulation must be dependent on the other events. This dependence is mimicked in the simulation by so-called density fields. These density fields determine the number density of each particle specie during the simulation. These fields are then used to determine the MFP and in the selection of the interaction type.

\begin{figure}[tbp]
  \centering
  \includegraphics[width=0.8\textwidth]{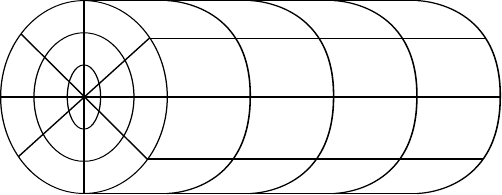}
  \caption{Schematic representation of the source segmentation for density fields and particle current calculation. The source is segmented into evenly spaced segments in longitudinal, radial and azimuthal direction predefined by the user. The segment boundaries are used as predefined planes for the determination of particle currents.}
  \label{fig:segmentation}
\end{figure}

In the simulation, the source is \changed{split} into evenly spaced intervals in longitudinal, radial and azimuthal direction, predefined by the user, see Figure~\ref{fig:segmentation} for a schematic representation \changed{of the segmentation}. In each iteration step of the particle list, it is recorded how long a particle with index $i$ stays in one of the  segments, labeled by the time $t_i$. The number density $n_\alpha$ of a species $\alpha$ is then proportional to the total time of all particles of the corresponding species spent in the segment $\sum_i t_i$ and the volume $V$ of the segment. In case of recombination, an additional term has to be added corresponding to the number of recombination $N_\gamma$ of the recombination partner $\gamma$ in the segment, resulting in
\begin{equation}
  n_\alpha = \frac{\sum_i t_i}{t_\text{sim} V} - \frac{N_\gamma}{V} \commaeq
  \label{eq:target_field_density}
\end{equation}
where $t_\text{sim}$ is called the simulated time \citep{Kellerer_2022}. It describes the simulated physical time that has passed after simulating a fixed number of primary events $N_\text{prim.}$. In the case of radioactive beta decay at KATRIN, it can be directly determined by the activity $a$ through
\begin{equation}
  t_\text{sim} = \frac{N_\text{prim.}}{a}\, . \label{eq:simulated_time}
\end{equation}
The activity of the source is not dependent on the interaction of the charged particles in the source, and can therefore be evaluated from the predefined amount of neutral gas in the source. Furthermore, it is assumed that the neutral gas flow is independent of the charged particle interactions. This is valid if the number of charged particles is negligible with respect to the number of neutral ones.

\subsection{Current collection}
\label{ssec:current_collection}

The second main objective of the code is to determine the particle currents in the source. This task is performed through so-called virtual barriers: predefined planes within the source, \changed{which register particle crossings.} The total current through the segment faces is then given by \citep{Kellerer_2022}
\begin{equation}
  j = \frac{N^+ - N^-}{A \, t_\text{sim}} \commaeq
\end{equation}
where $N^+$ and $N^-$ are the number of particles crossing the faces in positive and negative direction respectively, $A$ is the area of the face and $t_\text{sim}$ the simulated time, see equation~\eqref{eq:simulated_time}. For convenience, the location of the virtual barriers are chosen to coincide with the segment boundaries, see Figure~\ref{fig:segmentation}, of the density field calculation.%, see section~\ref{ssec:denstiy_fields}.

\section{Description of the implementation}
\label{sec:implementation}

For the simulation, we decided to use C++ as a programming language for several reasons. First, it can be used for object oriented programming. This makes the code easier to read, because the classes were closely modelled after their physical representation. Secondly, the accuracy of the simulation relies on high statistics. Thus, many particle decays need to be simulated, which requires a fast calculation. For the same reason, we decided to employ parallel calculation, see section~\ref{ssec:parallelization}.

There are three basic building blocks of the simulation. First, the simulation itself is treated as a class, which is responsible for initialization of the simulation environment and guiding through the flow of the simulation, see Figure~\ref{fig:simulation_flow}. Second, the particle class, which represents the simulated physical particles through their physical position, velocity and type of specie. Third, the specie class, which encapsulates the general behavior of each particle type (\changed{e.g.} electrons) in the simulation.

The main feature of the species class is to provide access to the density, velocity distribution and drift velocity of a specific type of particle. All three quantities are used for the calculation of the MFP and the choice of the interaction partner. However, we decided to distinguish between two types of species, those who are tracked throughout the simulation (electrons and ions) and those who act as background (tritium gas). This way, it is more clear which specie fulfills which function, especially in the initialization process.
% Additionally, the particle class only accepts the tracked specie class as input, reducing programming errors in the development of new injectors or interactions.

The tracked specie class is composed of different classes, providing access to the behavior of each particle type. Hence, each tracked specie can be tailored to the requirements of the user, either for testing, but also for the inclusion of further particles types. Along these lines, the code can be adapted to other experimental environments, provided the knowledge of the interactions and their cross sections. Each of the class members fulfils a special task, which will be described in detail in the following sections.

\subsection{Movement}
\label{ssec:movement}

The movement of particles is outsourced to the mover class. It moves the particle dependent on the E/B fields at \changed{its} position, given the integrated distance a particle will move, see equation~\eqref{eq:distance_from_mfp}. The movement itself is divided into smaller steps. This way, it can be evaluated if the particle has left the simulation domain between the steps, which stops the movement. Apart from the simulation boundary, the movement is stopped if the particle leaves a segment or energy bin of the density fields. %, see section~\ref{ssec:denstiy_fields}. 
\changed{In this case} no interaction will be performed, but a new distance will be drawn from the MFP at the new position of the particle, \changed{according to} equation~\ref{eq:distance_from_mfp}. This procedure does not increase the mean distance a particle travels, if the traversed density stays the same. However, large gradients will be taken into account in the calculation. Hence, the distance a particle moves is not over- or underestimated when moving into higher or lower density regions respectively. However, the density fields need \changed{a higher resolution than the scale length of the density difference}, which increases the simulation time. Thus, the user needs to evaluate the trade off between simulation time and accuracy.

Stopping the movement if a particle has left a density field segment has an additional benefit: it can be easily evaluated how long a particle has stayed in the segment. Each movement step the particle is alive, a counter variable is increased by the time the movement step took. If the movement is stopped this counter variable can be added to the corresponding density field segment.

Two different kinds of movers are implemented (Boris, drift), which differ in their calculation of the new position and velocity. In general, the Boris mover provides a high accuracy, because it resolves the gyro motion of the particles, but it therefore needs more calculation time. Vice versa, the drift mover provides a fast calculation time, because it does not resolve the gyro motion, but it is therefore less accurate. Both movers will be presented shortly in the following.

\paragraph{Boris mover}

The Boris mover uses the Boris push algorithm \citep{Penn.2003}. \changed{It solves the coupled differential equations
\begin{align}
  \frac{x_{i+1} - x_i}{\Delta t} &= v_{i+1/2} \comma \\
  m \frac{v_{i+1/2} - v_{i-1/2}}{\Delta t} &= F(v_i) \comma
\end{align}
where the index $i$ denotes the timestep at which the value is available. $F$ is the force acting on the particle, in this case the Lorentz force. The position update is straight forward and does not need any more explanation. The velocity update is more complicated, because the force on the particle is dependent on the velocity of the particle. By assuming that the velocity between two timesteps can be calculated as the mean of the values, it can be formulated that the velocity of the next timestep must read 
\begin{align}
  \bold{v}_{i+1/2} &= \bold{v}_{i-1/2} + \frac{q \Delta t}{m} \cdot \left(\bold{E}_i + (\bold{v}_{i+1/2} + \bold{v}_{i-1/2}) \times \bold{B}_i \right) \fullstop
\end{align}
}

\changed{
The change of the velocity vector is calculated in three steps. First, the particle is accelerated by half a timestep through the electric field
        \begin{equation}
            \bold{v}^- = \bold{v}_{i-1/2} + \frac{q \Delta t}{2 m} \bold{E}_{i-1/2} \fullstop
            \label{eq:boris_push_start}
        \end{equation}
        Secondly, the particle is accelerated by the magnetic field. This can be described through
        \begin{eqnarray}
            \bold{\Omega} =  \frac{q \Delta t}{2 m} \bold{B}_i \comma \\
            \vec{t} = \frac{2 \bold{\Omega}}{1 + \bold{\Omega} \cdot \bold{\Omega}} \comma \\
            \bold{v}^\prime = \bold{v}^- + \bold{v}^- \times \bold{\Omega} \\
            \bold{v}^+ = \bold{v}^- + \bold{v}^\prime \times \bold{t} \fullstop
        \end{eqnarray}
        Thirdly, the particle is accelerated again by half a timestep through the electric field
        \begin{equation}
            \bold{v}_{i+1/2} = \bold{v}^+ + \frac{q \Delta t}{2 m} \bold{E}_{i-1/2} \fullstop
            \label{eq:boris_push_end}
        \end{equation}
In total, the new velocity was derived from the mass and charge of the particle and the extrapolated electromagnetic field values. }

%for the movement step described in the previous paragraph (see \citep{Kellerer2022_1000143868} for a in depth description of the Boris push algorithm).

However, the steps are not chosen by a given length but a given timestep $\Delta t$, which is needed as an input for the Boris push algorithm. The moved distance is calculated after the movement \changed{step}. This  distance is then subtracted from the total distance the particle has still to move. When this distance is smaller than zero the movement will be ended.

The timestep has to be provided by the user. \changed{In order to resolve the gyro-motion of all particles of a species, the timestep should be chosen depending on the highest possible gyro frequency $\Omega_{g,\text{max}}$ of that specie. Thus }
%It should be based on the highest gyro frequency possible in the simulation domain $\Omega_{g,\text{max}}$ and fulfil
\begin{equation}
\label{eq:boris_timestep_requirement}
  \Delta t < \frac{1}{\Omega_{g,\text{max}}} 
\end{equation}
\changed{should be fulfilled. Additionally one has to consider that the maximal distance moved during one timestep is small enough that no segments might be skipped.}
This timestep is also used for the determination of the time a particle spends in a density field segment, see previous section.

The Boris push algorithm was chosen, because it is numerically stable with high precision and efficiency \citep{Kilian.2015}. The method can be used for low relativistic particles, up to a gamma factor of $\gamma \approx 1000$ \citep{Vay.2008}. At the KATRIN experiment, only electrons of energies up to \SI{32}{keV} (Krypton conversion electrons) are created. This results in a gamma factor of $\gamma = 1.06$, which is well below the limit of the Boris push.

\paragraph{Drift mover}

The drift mover is based on the description of the movement of the particle around its gyrocenter and the drift movement of the gyrocenter, see \citep{Chen.1984}. In the current version of the code, the drift mover should only be used for a constant magnetic field. No magnetic field gradients are considered, because the magnetic field at KATRIN is constant over most parts of the source \citep{Kellerer2022_1000143868}.

In the drift mover, the movement is divided into smaller steps \changed{of length $\Delta x$}, given by the smaller value of the maximal step length $d_\text{max}$ (provided by the user) and the leftover distance the particle has still to move. \changed{The distance $d_\text{max}$ should be chosen smaller than the smallest extent of a segment. Otherwise time spent in a new segment might not be correctly counted or certain segments might even be skipped completely.} The movement is ended, when the leftover distance is zero. Each movement step \changed{is} subdivided into the change of the particle position and the change of the velocity, both described in the following.

The position change is described by the movement around the gyrocenter $\Delta \mathbf{x}_\text{c}$, the movement of the gyrocenter parallel to the magnetic field $\Delta \mathbf{x}_\parallel$ and the drift movement of the gyrocenter $\Delta \mathbf{x}_\text{D}$.

The drift movement of the gyrocenter is calculated from the $E \times B$ drift velocity
\begin{equation}
  \mathbf{v}_\text{D} = \frac{\mathbf{E}\times \mathbf{B}}{\abs{\mathbf{B}}^2}
\end{equation}
and the timestep length $\Delta t$
\begin{equation}
  \Delta t = \frac{\Delta x}{\sqrt{\abs{\mathbf{v}_\parallel}^2 + \abs{\mathbf{v}_\perp}^2 + \abs{\mathbf{v}_\text{D}}^2}} \commaeq
\end{equation}
where $\Delta x$ is the step length of the movement step and $\mathbf{v}_\parallel$ and $\mathbf{v}_\perp$ the velocity parallel and perpendicular to the magnetic field. The moved distance through the drift motion evaluates to
\begin{equation}
  \Delta \mathbf{x}_\text{D} = \Delta t \cdot \mathbf{v}_\text{D} \fullstop
\end{equation}

The movement of the gyrocenter parallel to the magnetic field is calculated from
\begin{equation}
  \Delta \mathbf{x}_\parallel = \Delta t \cdot \mathbf{v}_\parallel \fullstop
\end{equation}

The position change around the gyrocenter is calculated in three steps. First, the gyro radius is determined
\begin{equation}
  r_\text{g} = \frac{m \abs{\mathbf{v}_\perp}}{q \abs{\mathbf{B}}} \commaeq
\end{equation}
from the magnitude of the perpendicular velocity $\abs{\mathbf{v}_\perp}$, the particle mass $m$ and charge $q$, which in turn is then used to determine the vector from the particle to the  gyrocenter $\mathbf{x}_\text{pc}$
\begin{equation}
  \mathbf{x}_\text{pc} = - r_\text{g} \frac{\mathbf{v}_\parallel \times \mathbf{v}_\perp}{\abs{\mathbf{v}_\parallel \times \mathbf{v}_\perp}} \fullstop
\end{equation}
Second, the gyro frequency is calculated through
\begin{equation}
  \omega_\text{g} = \frac{q \cdot \abs{\mathbf{B}}}{m} \fullstop
\end{equation}
The gyro frequency is then used to calculate the relative pitch angle change towards the gyrocenter
\begin{equation}
  \Delta \theta = \Delta t \cdot \omega_\text{g} \fullstop
\end{equation}
Third, the movement of the gyrocenter is calculated by the rotation of $\mathbf{x}_\text{pc}$ around $\mathbf{v}_\parallel$ with the angle of rotation $\Delta \theta$ through Rodriges' rotation formula \citep{10.5555/561828}. In total, the position changes calculates to
\begin{equation}
  \Delta \mathbf{x} = \Delta \mathbf{x}_\text{c} + \Delta \mathbf{x}_\parallel + \Delta \mathbf{x}_\text{D} \fullstop
\end{equation}

The velocity after movement is determined by the acceleration of the particle through the electric field $\Delta \mathbf{v}_\text{E}$, through the rotation of the perpendicular velocity around the gyrocenter $\mathbf{v}_{\perp,\text{rot}}$ and the initial parallel velocity
\begin{equation}
  \Delta \mathbf{v} = \mathbf{v}_\parallel + \Delta \mathbf{v}_\text{E} + \mathbf{v}_{\perp,\text{rot}} \fullstop
\end{equation}

The acceleration through the electric fields calculates to
\begin{equation}
  \Delta \mathbf{v}_\text{E} = \Delta t \frac{q \mathbf{E}_\parallel}{m} \commaeq
\end{equation}
where $\mathbf{E}_\parallel$ is the electric field parallel to the magnetic field.

The rotation of the perpendicular velocity is again calculated using Rodriges' rotation formula with the axis of rotation $\mathbf{v}_\parallel$ and angle of rotation $\Delta \theta$.

\changed{
\paragraph{Comparison}
The choice of which mover algorithm with which parameters should be used for which particle specie is a complex one. An estimate for when both algorithms should take the same amount effort can be made by comparing the distances moved by the particle during one movement step. If the distance for the Boris mover $d_\text{Boris}$ is larger than the user-defined distance $d_\text{Drift}$, then the Boris version has a better performance.
The distance $d_\text{Boris}$ can be estimated using the timestep $\Delta t$ and the typical velocity $v_{typ}$ of the considered particle species through
\begin{equation}
\label{eq:boris_distance_estimate}
d_\text{Boris} \approx v_{typ}\Delta t \overset{Eq.\, \eqref{eq:boris_timestep_requirement}}{<} \frac{v_{typ}}{\Omega_{g, \text{max}}} \fullstop
\end{equation}
Using the requirement $d_\text{Drift} < d_\text{Boris}$ and the formula for the gyro-frequency one arrives at
\begin{equation}
\label{eq:magneticField_req_boris}
B_\text{max} < \frac{v_{typ} m}{q d_\text{Drift}} \commaeq
\end{equation}
as a threshold for the magnetic field for which the Boris mover outperforms the Drift mover. For electrons originating from a beta decay (approximately \SI{20}{\kilo eV}) and $d_\text{Drift} = \SI{1}{\mm}$ this leads to a value of $B_\text{max} \approx \SI{470}{\milli\tesla}$. For a thermal population of T$_2^+$ at \SI{80}{\kelvin} with the same value for $d_\text{Drift}$ the value is $B_\text{max} \approx \SI{21}{\milli\tesla}$, and for thermal electrons the value is even lower. However, the timestep also needs to consider the same restrictions as the choice of $d_\text{Drift}$, meaning that the timestep also has to be chosen such that no segments might be skipped during one movement step. This in turn means that the Drift and Boris mover have at best the same computational effort, even at low magnetic fields.\\
If no electric fields are present and the magnetic field is uniform everywhere the above discussion is the only criterion for the choice of the mover. Therefore in this case only the Drift mover should be chosen since it is always at least as performant as the Boris one and more importantly the drift approximation becomes exact.\\
However if the magnetic field is not uniform or zero the Boris mover has to be used since the Drift mover has no support for gradients in the magnitude of the magnetic field.\\
If the magnetic field is uniform and electric fields are present, the choice becomes more complex. For very strong magnetic fields, i.e. a slow Drift motion and a tight gyro-orbit the Drift mover is still a good approximation and also outperforms the Boris mover. The opposite is the case for very weak magnetic fields, the approximation becomes bad and the Boris mover should be used instead. 
% probably write down the actual requirements for the drift approximation and maybe even cut the first discussion completely
}

\subsection{Termination}
\label{ssec:termination}

The termination of particles is addressed by the terminator class. It checks for three different conditions. First it checks, if the particle has left the simulation geometry. For the KATRIN experiment the geometry can be approximated by a cylinder. However, the radius and height have be provided by the user for special analysis. Second, the terminator class checks if the particle has exceeded the user defined number of maximal interactions. This counter was added, to ensure that no trapped particles are tracked by the simulation. Otherwise the simulation would not end. \changed{To assure that the simulation continues to be physically accurate, it is important to choose an appropriate value for the number of maximal interactions. A discussion of this problem will be given below in the test cases section~\ref{sec:test}.} This trapping can occur for example through opposite neutral gas flow and reflection by electric fields. Third, the terminator class checks if the particle has exceeded the user defined lifetime. This counter ensures, that particles are not stored indefinitely. This could occur through special configuration of magnetic and electric fields.

In addition to the terminator class, the user can choose to add a deletion logger to the specie class. This class is evoked if the particle is deleted. It stores information on the type of deletion, namely if the particle has hit a wall, recombined, formed a cluster, or exceeded the maximum time alive or maximum number of interactions. These values are also added to the output of the simulation. So, this information can be used for debugging, but it can also be used for physical results. For example, the number of recombinations provides an insight if recombination is a dominant process.

\subsection{Interactions}
\label{ssec:interactions}

\begin{figure}
\centering
\includegraphics[width=.9\textwidth]{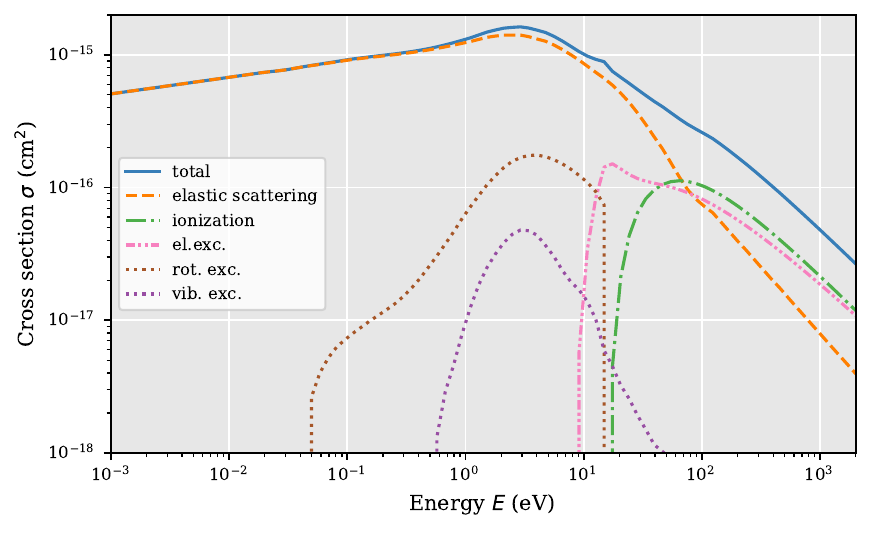}
\caption{Cross sections for electron and H$_2$ interactions. The excitation cross sections depicted are the sum over the different excitation states. Image from \citep{Kellerer2022_1000143868}.}
\label{fig:et2_interactions}
\end{figure}

\begin{figure}
\centering
\includegraphics[width=.9\textwidth]{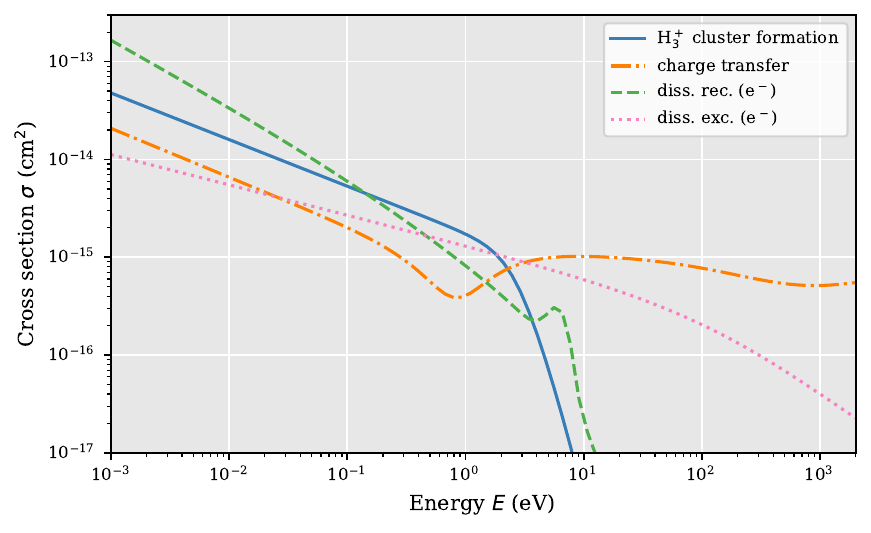}
\caption{Cross sections for H$_2^+$ interactions with H$_2$. The dissociative recombination and excitation cross sections are in relation to the kinetic energy of the electrons in the rest frame of the H$_2^+$ ions. Image from \citep{Kellerer2022_1000143868}.}
\label{fig:t2p_interactions}
\end{figure}

\begin{figure}
\centering
\includegraphics[width=.9\textwidth]{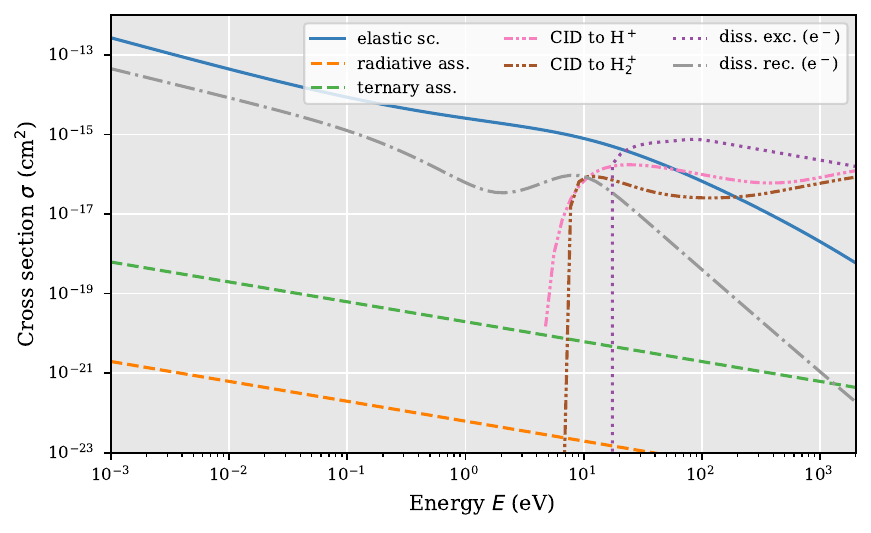}
\caption{Cross sections for the formation and destruction of H$_3^+$ within a H$_2$ environment. The dissociative recombination and excitation cross sections are in relation to the kinetic energy of the electrons in the rest frame of the H$_2^+$ ions. Image from \citep{Kellerer2022_1000143868}.}
\label{fig:t3p_interactions}
\end{figure}

Interactions of particles are processed in the interaction class. Additionally, it is used to provide access to the cross section of the interaction and the \changed{target-density scaled} cross section, which \changed{is} used in the calculation of the MFP, \changed{see equation \eqref{eq:mean_free_path}}, and in the selection of the interaction. %, see section~\ref{ssec:general_diagram}. 
The interaction class itself is an interface class. The details of the interaction are implemented in derived classes. This way it is easy to add more types of interactions in the future. All interactions are initiated using a given cross section and target specie. For the cross section, we decided to use a one dimensional functional. This way the user can either specify an implemented cross section, or provide an own function in the configuration file, see section~\ref{ssec:initialization}. The target specie can either be a tracked or background specie, \changed{with some restrictions dependent on the specific interaction}. 
%This allows for a high flexibility of the interaction class.

Currently there are seven types of interactions implemented: elastic scattering, clustering, recombination, free movement, charge exchange, excitation and ionization. In general, all types can be added to all species. However, the user has to decide if the specific use is sensible. In the following, the implementation of each interaction will be described. For the physical explanation, please refer to \citep{Kellerer_2022}.

\paragraph{Elastic scattering}
\changed{
The most prominent interaction between low energy particles is elastic scattering, in which the total kinetic energy is conserved. For electrons colliding with molecular tritium this interaction can be written as
\begin{align}
\label{eq:elastic_scattering}
\begin{split}
e^- + \text{H}_2 &\rightarrow e^- + \text{H}_2 \fullstop
\end{split}
\end{align}
This interaction can be used for any tracked specie colliding with a background species as the target.
% Note: a tracked species can also be used as the target but in that case both a drift velocity and a velocity distribution have to be specified.
The values of the cross sections for electrons can be seen in Figure~\ref{fig:et2_interactions} and for H$_2^+$ in Figure~\ref{fig:t2p_interactions}.}

The elastic scattering class is initiated using the cross section, the target species and a scattering angle distribution. This distribution \changed{is} used to draw a scattering angle, given the energy of the incoming particle. \changed{The available options are} an isotropic distribution, a distribution provided by a file and a \changed{composition} of distributions.

During evaluation of the scattering angle, the energy bin is determined and the corresponding angle distribution selected. The scattering angle is then drawn from \changed{this} angle distribution \changed{using a piecewise linear distribution}. 

%Note: Somewhere the structure for input files has to be explained! -> explain in README, short desc. in text here

In some cases, there is not enough data from one source alone, to cover the whole impact energy range. This is the case for the scattering of electrons with tritium molecules, \changed{where data for the cross section was only available for low impact energies}. At high energies, the scattering angle distribution is determined from the external simulation Elsepa \citep{Salvat:2005aa}, at lower energies isotropic scattering can be assumed. \changed{For these cases, the scattering angle distribution can be composed using the known distributions and a threshold energy defining point of transition between the distributions.}
%For these cases, the angle can be determined through the composition of two distributions.

The cross section of elastic scattering was measured at different experiments in the energy range of \changed{\SI{2e-2}{eV} to \SI{100}{eV}}. The results of these experiments were combined by \citep{Tawara.1990} and parameterized by \citep{Yoon.2008}.

The interaction itself is performed in multiple steps. First, the velocity of a target particle is determined. It is \changed{a} compound of the drift velocity of the species and a velocity, which is drawn from the underlying velocity distribution. For the tritium gas, the drift velocity is determined in external simulation and the velocity distribution is given by a Maxwell-Boltzmann distribution \citep{Kellerer_2022}.
%Note: the following sentence should be a lie. A tracked species should throw an error if any of the two components is initialized with "null"...
%For tracked particles the drift velocity is assumed to be zero and the velocity distribution is given by the density field.
The velocity of the target particle is used to determine the impact energy in the \changed{center of mass} frame. This energy is then used for the evaluation of the polar scattering angle $\theta$. A second scattering angle is drawn in the interval $[0, 2\pi)$, responsible for the change of momentum in azimuthal direction. Both angles and the target velocity are then used to rotate the velocity vector of the impact particle in the center of mass frame, see \citep{Kellerer_2022} for the full calculation. This rotated velocity is then used as the new velocity of the impact particle.

\paragraph{Clustering}
\changed{
There are currently two types of clustering reactions implemented. The first one is a two particle process:
\begin{align}
\label{eq:clustering_radiative}
\begin{split}
\text{H}^+ + \text{H}_2 &\rightarrow \text{H}_3^+ + h\nu \quad \text{or} \\
\text{H}_2^+ + \text{H}_2 &\rightarrow \text{H}_3^+ + \text{H} \fullstop
\end{split}
\end{align}
}

\changed{
The second type of clustering includes three interaction partners:
\begin{equation}
\label{eq:clustering_tertiary}
\text{H}^+ + \text{H}_2 + \text{H}_2 \rightarrow \text{H}_3^+ + \text{H}_2 \fullstop
\end{equation}
Since for both types of reactions only the rate coefficients from \citep{Gerlich.1992,Paul.1995} are available (which have been determined from an experiment), the cross section itself has to be approximated. This has been explained in \citep{Kellerer2022_1000143868} and the resulting cross sections are shown in Figure~\ref{fig:t3p_interactions}.
%Note: if i'm honest, way to little explanation, since rate coefficients themselves are only at a specific temperature etc...
}
The clustering class is initiated using these \changed{approximated} cross sections, the target specie and the specie, which is the result of the clustering process. In the interaction itself it is assumed that the new particle follows the direction of movement of the target particle, but with different absolute velocity. The new absolute velocity is calculated from the velocity of a target particle and the new particles mass, since the initial direction is less important due to a high probability of collisions with the neutral gas, which quickly randomizes the velocity direction. The velocity of a target particle is determined \changed{analogous} to its calculation in elastic scattering.

\paragraph{Recombination}
\changed{
Recombination is the interaction in which an ion species and an electron species form a neutral particle. Only two body processes are considered for this case:
\begin{equation}
\label{eq:recombination}
\text{H}^+ + e^- \rightarrow \text{H} + h\nu \fullstop
\end{equation}
}
The recombination class is initiated using the cross section and the target specie. The interaction itself has only two steps. First, the target density field logs one deletion at the corresponding position, see section~\ref{ssec:denstiy_fields}. Second, the impact particle is marked for deletion. The different recombination cross sections are given by analytic calculations \citep{Janev.2003,Kotelnikov.2019,Pettersson.2015} \changed{and are plotted in Figures~\ref{fig:t2p_interactions} and \ref{fig:t3p_interactions}.}

\paragraph{Free movement}
The free movement interaction is only a helper class, which is necessary if no other interaction should be used \changed{or if the cross section is zero, e.g. if the energy is below the threshold of an interaction}. It provides the simulation with a very small cross section, which results in a very large mean free path. This large \changed{MFP} guarantees a\changed{n interaction-}free movement through the simulation domain. If the interaction step is evoked for a particle nothing is done.

\paragraph{Charge exchange}
\changed{
Only the following type of charge exchange interactions is considered 
\begin{equation}
\label{eq:charge_exchange}
\text{H}_2^+ + \text{H}_2 \rightarrow \text{H}_2 + \text{H}_2^+ \commaeq
\end{equation}
i.e. charge exchange processes are only possible between a background species and the corresponding singly charged ion.
}
The charge exchange class is initiated using the cross section and the target specie. No product specie is necessary, because product and impact specie are the same. In the interaction, only the velocity of a target particle is determined, analogue its calculation in elastic scattering. This velocity is then used as the new velocity of the impact particle. Cross sections for this process class can be found in \citep{Janev.2003} \changed{ and are plotted in Figure~\ref{fig:t2p_interactions}}.

\paragraph{Excitation}
\changed{
The target particle can be excited if the collision energy is above the corresponding excitation energy. This is mainly the case in electron collisions with the background gas
\begin{equation}
\label{eq:excitation}
e^- + \text{H}_2 \rightarrow e^- + \text{H}_2^* \commaeq
\end{equation}
where H$_2^*$ is either rotationally, vibrationally or electronically excited version of H$_2$.
}
The excitation class is initiated using the cross section, the target specie and the excitation energy. In the interaction itself, first the velocity and direction of movement of the impact particle is determined. Second the kinetic energy of the particle \changed{is} calculated and reduced by the excitation energy. Lastly, the new velocity is determined from the reduced energy and direction of movement. This is a simple but effective method for excitation processes. However, no scattering is considered here. This is justifiable, because excitation processes are often much \changed{less }likely than elastic scattering processes and therefore scattering angle effects do not play a significant role. The excellent paper by \citep{Janev.2003} provides more insight into the cross sections. \changed{The cross sections are plotted in Figure~\ref{fig:et2_interactions}, where a sum over all final excitation states has been performed}.

\paragraph{Ionization}
\changed{
In a molecular background gas there are two types of electron ionization processes. In the first one, the target only loses an electron
\begin{equation}
\label{eq:ionization_non_diss}
e^- + \text{H}_2 \rightarrow e^- + e^- + \text{H}_2^+ \fullstop
\end{equation}
In the second one the target itself also dissociates:
\begin{equation}
\label{eq:ionization_diss}
e^- + \text{H}_2 \rightarrow e^- + e^- + \text{H}^+ + \text{H} \commaeq
\end{equation}
where the H might also be left in an excited state.
}
The ionization class is initiated using the cross section, the target specie, the new electron specie, the new ion specie, the ionization energy, \changed{the ionization energy distribution} and the scattering angle distribution. It is developed for ionization of the target specie by impact of a particle, which is much lighter than the target. However, the code does not check for this condition.

In the interaction, first the velocity of a target particle is determined \changed{analogous} to its calculation in elastic scattering. Then, the impact energy in the \changed{center of mass} frame $E_\text{i}$ is calculated. If the impact energy is below the ionization energy $E_\text{ion}$, no ionization could occur and a new target particle velocity is drawn. This procedure can take a long time. However, the cross section of ionization decreases significantly \changed{close to} the ionization energy. Thus, very few particles with energies close to the ionization energy will undergo ionization, which will therefore not increase the simulation time significantly. \changed{Once} a suitable target particle velocity is determined, the impact energy is used in the ionization energy distribution
% Note: should probably be plotted...
% Note: mention how the particles are actually differentiated? (fast->impact, slow->ejected)
 to draw the impact energy after collision $E_\text{i}^\text{new}$. The energy of the new electron $E_\text{e}$, also called ejected electron in the following, then follows \changed{from}
\begin{equation}
  E_\text{e} = E_\text{i} - E_\text{ion} - E_\text{i}^\text{new} \fullstop
\end{equation}
Next, the direction of movement has to be determined. For both the impact particle and the ejected electron, the direction of movement is calculated as a rotation of the impact velocity direction in the \changed{center of mass} frame by the polar and azimuthal angle. The azimuthal angle of the impact particle $\phi_\text{i}$ is drawn \changed{uniformly} from the interval $[0,2\pi)$. The direction of the \changed{ejected} electron is set to the opposite direction. Thus, the azimuthal angle of the ejected electron $\phi_\text{e}$ is calculated from
\begin{equation}
  \phi_\text{e} = \phi_\text{i} - \pi \fullstop
\end{equation}
The polar angle of both particles is calculated, using the derivation of Grosswendt and Waibel \citep{Grosswendt:1978aa}. Lastly, the velocity of the ejected ion is set to the velocity of the target particles, assuming no recoil of the target particle in the interaction. As described before, this assumes a large mass difference of the impact particle and the target particle, which is the case for the dominant process in the source: ionization of tritium gas by electron impact.

The cross sections and more importantly the distribution of secondary particles are provided by \citep{EugeneRudd.1989,Kim.1994}. \changed{The total ionization cross section is also shown in Figure~\ref{fig:et2_interactions}.}

\subsection{Calculation of fields}
\label{ssec:field_calculation}

The density fields are implemented as a class. Each specie holds its own member of this class, which simplifies the evaluation of the fields during the simulation, for example for the calculation of the MFP.
The class \changed{stores} the time particles of the species have spent in a segment of the source, the number of recombination in this segment and the derived density itself. They are represented as four dimensional arrays. The fastest axis (third axis) is used for the different energy bins. The fastest axis refers to the axis for which the entry can be accessed the quickest. \changed{This is an effect created by caching. Since data blocks located closely in the memory are pre-loaded into the cache, which reduces the access time.} This reduces the calculation time of the determination of the density in a segment, which is mainly a sum over all energy bins. It is evoked multiple times during the particle update, which requires a fast calculation. The use of the other axes follows the consideration of how often the bin changes in longitudinal, azimuthal and radial direction. The magnetic field is most commonly aligned in longitudinal direction. Thus, the particles mostly travel in longitudinal direction. Therefore, the second fastest axis is used for the longitudinal direction. The other two directions are similar in their usage. Because of the magnetron drift in radial inhomogeneous magnetic fields, it was decided to use the azimuthal direction next, followed by the radial direction.

The size of the arrays is derived from the user input. It is calculated from the desired number of segments and a built in under- and overflow. This way, if a particle can not be classified into one of the segments, the information is still recorded in the simulation. The energy bins are created in logarithmic scale with one bin each for under- and overflow. It can be set differently for each specie. Thus, important effects for each specie can be resolved more easily. In longitudinal direction, the bins are divided evenly with one bin each for under- and overflow. In general, these under- and overflow bins should not be filled during the simulation, because this would mean that a particle has left the simulation domain and is still alive. Hence, they can be used for debugging purposes. In radial direction, there is only an overflow bin, because the radius is lower bound to zero. In azimuthal direction no over- and underflow bins are used, due to the circular nature of the problem.

The calculation of the density also needs the physical time, see equation~\eqref{eq:target_field_density}. This time is provided by a pointer to a timer class. This class is called, when new particles are injected in the simulation domain, see section~\ref{ssec:injection}, \changed{to increase the physical time}.

The current collector is structured similar to the density field class. The difference being, that it holds the crossings in positive and negative direction through the predefined faces in three dimensional arrays for crossings in longitudinal, azimuthal and radial direction instead. The faces can be aligned \changed{with} the segmentation of the density fields, but can in general be set \changed{arbitrarily}.

% Next sentence is a lie, only the binning of the density field is used, the current collector does not even have the functions to perform the corresponding checks
%However, the mover uses both the density fields and current collectors to evaluate possible endpoints of movement, see section~\ref{ssec:movement}. Thus, a different segmentation will lead to a longer simulation time.

\subsection{Injection}
\label{ssec:injection}

The injection of new \changed{decay} particles to the particle list is managed by the injector class. It is initiated by providing the event position distribution, the mother drift velocity and a list of species, with their energy and injection angle distribution. The event position distribution is used to determine the position, at which the new particles are to be injected. For each injected specie, the kinetic energy is drawn from the energy distribution. The corresponding velocity vector is drawn from the injection angle distribution and scaled by the kinetic energy. This velocity is then boosted by the mother drift velocity, which represents the movement of the decaying gas.

In the case of the KATRIN source, the injected species are electrons, with an isotropic injection angle distribution and a Fermi energy distribution \citep{Fermi:1934aa} and T$_2^+$ ions, also with an isotropic injection angle distribution, but with a Maxwell-Boltzmann energy distribution. T$_2^+$ is used instead of HeT$^+$ due to the fact that both molecules have almost the same mass and no cross section data could be found in the literature for the latter. Nuclear recoil from the $\upbeta$-decay is not considered. This is again due to the higher interaction probability of the ions with the neutral gas, which thermalizes them more quickly. The mother drift velocity is provided by the drift velocity of the neutral tritium gas.

The injector class can also be used for the injection of a stream of directed particles. This is necessary, either for the evaluation of the influence of an electron gun in the experiment, or for test purposes.

After the injection of particles, the timer class is provided with the time, that has passed between injections. This time is based on the user defined activity. 
%% Might imply that the code can do that, which it can't 
%It can be calculated from the integrated density in the source and the decay constant.

\subsection{Parallelization}
\label{ssec:parallelization}

\changed{Even physical processes with a small cross section (and consequently low probability) will have an effect on the electron spectrum. In order to account for all processes high statistics is necessary, which may not be feasible on single core systems. Parallelization is therefore absolutely necessary.}
%In order to obtain reasonable and robust results of the simulation, high statistics is necessary. This is especially the case, if also improbable effects like the excitation by rotation, should be visible in the results. The goal of high statistics can be reached by parallel computation.
An asset of the code is, that the Monte Carlo events of each particle track can be dependent on the other, for example in the case of recombination. In order to obtain this strong point, the number of desired events per CPU core are further subdivided into cycles with a fixed length. After each cycle, a synchronization step between the CPU cores is initiated. During the synchronization step, the data on the density fields is shared among all CPU cores. Since the simulation is set up the same for all CPU cores, this step only needs an additive operation on the array of cumulated times%, see section~\ref{ssec:denstiy_fields}, 
and an additive step, to share the passed time. Additionally to the density field synchronization, also the data of the current borders and deletion loggers are synchronized, which is not strictly necessary, but keeps the data congruent on all CPU cores and makes the output step easier, see section~\ref{ssec:output}.

The benefit of this parallelization scheme lies in its simplicity. It is easy to include more CPU cores for the calculation. Only the synchronization step takes slightly longer. However, the main simulation effort lies in the movement and interaction step. Thus, the synchronization does not play a significant role. Furthermore, there is no need of a CPU core, which directs tasks to the other CPU cores. Hence, little communication and data exchange is necessary between CPU cores, which saves calculation time.

The simplicity of the method also produces some drawbacks. First, the load on the CPU cores can be different even though the number of events was set to the same value because of the creation of new particles through ionization. However, if the cycle length is chosen large enough, the load can balance out. The specific choice of cycle length depends on the simulated problem and has to be chosen by the user.

\begin{figure}[tbp]
  \centering
  \includegraphics[width=1.0\textwidth]{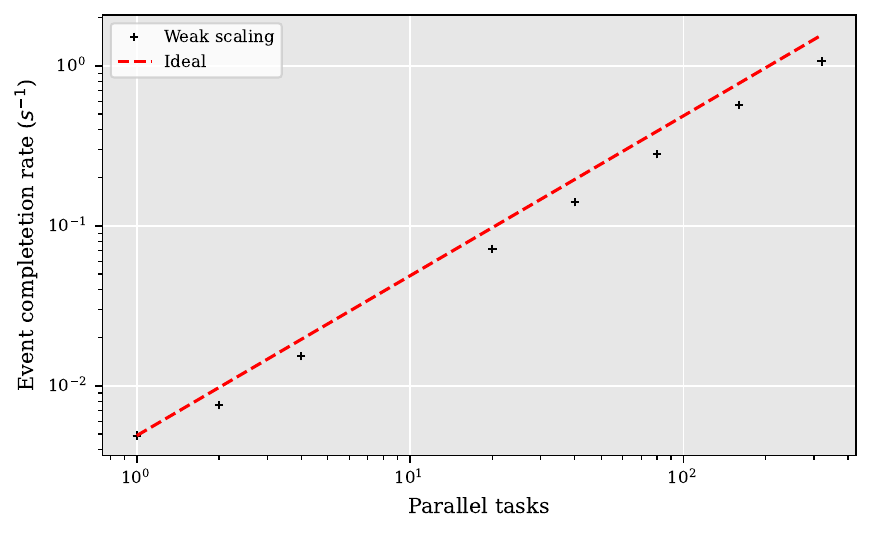}
  \caption{Weak scaling graph. Simulated using standard KATRIN parameters with increasing number of events. The load was \changed{kept constant on the} CPU cores. It was recorded how long the simulation took from start to finish, including output. The result of a single CPU core is used for prognosis of the ideal scaling. The simulations were performed on the bwForCluster NEMO\citep{nemo_cluster}.}
  \label{fig:scaling_graph}
\end{figure}

%The simulation code used for this study was parallelized to improve the efficiency of the computation, as described in the previous paragraph.
The effectiveness of this parallelization strategy can be seen in the scaling graph shown in Figure~\ref{fig:scaling_graph}. 
\changed{The graph shows the weak scaling behavior of the code, i.e. the number of events per core is kept constant and the number of cores is varied. In an ideal scenario the wallclock time needed until completion depends only on the number of events per CPU core, meaning that the number of events processed per second should increase linearly with the number of tasks. It can be seen that the graph follows this behavior quite well, except for the first data point, which is slower due to the initialization step that cannot be parallelized.}
%The graph illustrates the time it took for the simulation to complete for different numbers of events and CPU cores. As can be seen, the scaling is linear except for the first data point, which is slightly slower. This is due to the initialization of the simulation and the distribution of events to the different CPU cores. Once the simulation is properly initialized, the linear scaling is observed for all subsequent data points. 
This behavior is expected and is a common characteristic of parallelized simulations.

\subsection{Output}
\label{ssec:output}

The output of the simulation is written to a single HDF5 file \citep{hdf5}, which allows for extendable, structured data storage. Output is performed in two different modes. First, it is enacted after the initialization phase of the simulation to write non-changing data to the file for each of the tracked specie. This includes for example, the mass and charge of the species, the volumes and surface areas of the segmentation and the number of bins. Second, the output is performed to write the simulated data (from density fields, current collector, deletion logger) to file. This step is performed after \changed{an user-specified number of} CPU synchronization \changed{steps}, see section~\ref{ssec:parallelization}. In general, only the last output is of interest for the user. However, the steps in between can be used for debugging. Additionally, if the user has not set the correct wall time at the computing cluster an intermediate result is available.
\changed{Furthermore, the output of the different diagnostics (density field, current collector, deletion logger) can be modified to the specific use case. For each of the three fields the output interval can be set independently. The species for which the specific output should be performed and whether the latest output should be overwritten can be set individually as well.}
%Thus, the user can decide for each specie and output type (density field, current collector, deletion logger) the output interval and if the new data of the step should be saved separately or if the data of the old step should be overwritten. 
Overwriting saves disk space. Nevertheless, the size of the output file is comparably small, since it holds only derived properties; Depending of course on the number of segments chosen by the user. For most use cases the size of the output file is well below 1 gigabyte.

\subsection{Initialization}
\label{ssec:initialization}

The initialization of the simulation is administered by the simulation class.% using generators which are available for classes, requiring special user input.
The specific run instructions have to be provided in a JavaScript Object Notation (JSON) file. Due to its hierarchical structure, it is easy to read, test and extend by the user, even with little programming experience. We use the Jansson library \citep{jansson} to read in the file.

The JSON file is structured, such that each of the sections corresponds to one main idea of the code or class. Subsections are used, when the class is a composite of other classes. This way the parameter file stays extensible in the face of new methods and classes, which will be added in the future. A collection of JSON files for each \changed{major section} is provided together with the code for easy usage. A selection of \changed{the most important sections} will be discussed in the following.

\changed{
For input quantities that have a physical unit attached to them, e.g. the height and radius of the cylinder, the user can choose the units in the JSON file. During the parsing of the file an automatic conversion to SI units is performed. The conversion currently supports \si{mm}, \si{cm} and \si{m} for length units and \si{J} and \si{eV} for energy units.}

%If energy or length values are to be input by the user, the corresponding value has to be provided by its value and its unit. An automatic unit conversion will be taken care of by the program. Currently there are conversions available for \SI{}{cm}, \SI{}{mm} and \SI{}{m}, and for \SI{}{J} and \SI{}{eV}.

\paragraph{Functions}
%Functions are an integral part of the code, used for example to express cross sections. 
There are two types of functions implemented: a mapping of $\mathbb{R} \to \mathbb{R}$, called later on 1D-function or a mapping of $\mathbb{R}^3 \to \mathbb{R}$, called later on 3D-function.

\paragraph{Scalar field}
Scalar fields currently can only be initiated through one option, namely the \textit{ScalarFieldFunctionInZ}. As the name suggests, the return values only depend on the $z$-coordinate.
%it returns values, which are provided through a 3D-function. 
The general input of coordinates are transformed to cylinder coordinates $(r, \phi, z)$ and the corresponding function value is returned.

\paragraph{Vector field}
% Vector fields represent the idea, that for every point in space a certain three dimensional vector can be assigned.
A vector field can be initiated through three different options: \textit{VectorFieldConstant}, \textit{VectorFieldFunctionInZ} and \textit{VectorFieldFromArray}. %\changed{Any vector field object takes either a position as a Cartesian or cylinder vector and returns the value at that position as a cylinder vector.}
%For all three types, the general input of coordinates are transformed in the class to cylinder coordinates $(r, \phi, z)$ and the corresponding vector is calculated and returned.

\begin{itemize}
  \item \textit{VectorFieldConstant}: \changed{A constant cylinder vector is returned}%The return value of the vector field does not depend on its input and a constant vector is returned.
  \item \textit{VectorFieldFunctionInZ}: The returned vector only points in z direction. The norm of the vector is provided by a 3D-function.
  \item \textit{VectorFieldFromArray}: The returned vector is provided by precalculated values in arrays. The option \textit{radius} and \textit{height} is used together with the array size to define a cylindrical binning. %The input coordinate is transformed to a bin coordinate, which provides the index in the array. 
If the given coordinates is outside of the defined binning, a zero vector is returned.
\end{itemize}

\paragraph{Velocity distribution}
%The velocity distribution represents the idea, that a velocity vector can be sampled for each point in space.
Currently there is only one type of distribution implemented, the \textit{VelocityDistributionFromFunction}. However, in cases, where a velocity distribution is needed for a constructor but never will be used a `null' option is also available. For the \textit{VelocityDistributionFromFunction}, the velocity is sampled from a predefined function, the \textit{VelocityDistributionFunction}. Here, two options are available, the \textit{IsotropicDistribution} and the \textit{DirectedDistribution}. Both have in common, that first an energy is sampled from an given energy distribution. However, for the \textit{DirectedDistribution} this energy is used to scale a predefined direction vector using a predefined mass. For the \textit{IsotropicDistribution} the direction of movement is sampled from an isotropic distribution and then scaled to represent the correct energy. The mass can either be derived from a given particle type, or directly given as a \changed{arbitrary} value.

There are three different energy distributions available in the code, the electron energy distribution of tritium decay (\textit{TritiumBetaDecay}), the Maxwell Boltzmann distribution (\textit{MaxwellBoltzmann}) and a constant distribution (\textit{Constant}).
\begin{itemize}
  \item \textit{TritiumBetaDecay}: The energy is drawn using the known Fermi distribution \citep{Fermi:1934aa} and the rejection method.
  \item \textit{MaxwellBoltzmann}: The energy is drawn using the built-in gamma distribution of the C++ standard library and an user defined temperature.
  \item \textit{Constant}: A single energy is returned for all calls to the class.
\end{itemize}

\paragraph{Simulation}

In the simulation section, one can set the number of simulated events over all CPU cores (\textit{SimulatedEvents}). For KATRIN these events symbolize tritium decays. The events are distributed evenly on all CPUs. The parameter \textit{EventStride} is used to define the number of events per CPU after which the density field data is synchronized between the CPUs. %Also, output could be performed here.

\paragraph{Output}
In the output section, it is defined, where the output file is stored and with which filename.
\changed{The folder is interpreted relative to the directory the code has been called from.} %The folder is given relative to the home folder of the user or can be given as an absolute path. 
Additionally, the output specifics are defined in this section for each of the simulation data: density fields, current borders and deletion logger.

\begin{itemize}
  \item \textit{Interval}: Defines how often an output should occur \changed{relative to the CPU synchronization step. The value can either be set to \textit{EachSync} or a positive integer value}. %This can either be set to \textit{EachSync}, so each time the CPUs are synced the output is performed, or to a positive integer value. %This value specifies the rate of output. 
So the number two would correspond to an output every second CPU synchronization step.
  \item \textit{OverwriteLast}: Defines if each output should be stored on disk, or if each new output overwrites the last.
  \item \textit{Species}: Defines for which specie the output should be performed. It can either set to \textit{all}, or to a list of species given their specie name in the parameter file, eg. $[\text{electron}, \text{T2+}]$.
\end{itemize}

\paragraph{RandomNumbers}
In the random numbers section, the seed for the pseudo random numbers is set. It is used for the initialization of the random numbers of the first CPU. All following CPUs gain a seed given by the sum of the seed of the first CPU and its CPU rank. In the simulation, we use the 32 bit Mersenne Twister engine from the C++ standard library.

\paragraph{Species}

In the species section, all relevant information on the species used in the simulation is provided. It is subdivided into a section for the background species and tracked species. A specie is added to the simulation provided a unique key name and the necessary data, which is different for background species and tracked species. However, both sections need to specify the \textit{ParticleType}, which can either be `electron', `T2+', `T3+', `T+' or `T2'. This defines the mass and charge of the specie. Furthermore, background species need three sections to be initialized (\textit{ConstantDensityField}, \textit{VelocityDistribution} and \textit{DriftVelocity}), while tracked species need nine (\textit{Mover}, \textit{Terminator}, \textit{Reflector}, \textit{DensityField}, \textit{VelocityDistribution}, \textit{DriftVelocity}, \textit{Loggers}, \textit{CurrentCollector}, \textit{Interactions}).

\subsection{Particle list}
\label{ssec:particle_list}

The list used in the simulation to keep track of the simulated particles. Particles are added to the end of the list, when they are created through injection or by interactions like ionization. However, the next particle to simulate after the old one was deleted is not drawn from the front of the list but at random from all particles in the list. This decision was made, because the single events can be dependent on each other through the density fields and a bias towards injected particles should be avoided, especially because the injected particles have mostly higher energies than the others. Hence, it was decided to store the particles in a standard library vector \citep{Josuttis.2012}. To save computation time, the vector is initialized with a size of 100 elements, which marks a rough estimate of an upper limit on new particle per particle loop at standard KATRIN conditions.

\section{Test cases}
\label{sec:test}

The code was tested thoroughly, including a comparison of specific simulations with expectations. These expectations were formulated for scenarios where an analytical expression of the simulation output can be derived. For these simulations, each interaction was investigated detached from the others. \changed{More complex test cases, with the influence of multiple interactions at the same time were not performed due to the lack of analytical expectations or experimental data. \\
Within the `test\_cases' directory in the code repository a subdirectory for each test case is provided. These subdirectories contain the parameter and plotting files used to obtain the corresponding data presented in this section. The used output files will be provided via an external website.}
%The inclined reader is referred to \citep{Kellerer_2022} for a description of all the tests. Here, only a special test case shall be mentioned, because of its importance: the test of elastic scattering. One can show with this test that the method of density fields, the treatment of interactions in general and movement of particles are implemented correctly because they all are a prerequisite for the correct result.

\paragraph{Elastic scattering}
In this test case, monoenergetic particles are injected into a neutral gas reservoir with a Maxwell-Boltzmann distributed velocity. \changed{The temperature of the neutral gas was set to \SI{80}{\kelvin}. The electrons were injected in the middle of the simulation domain with an energy of \SI{8}{\milli eV} and a velocity direction in positive z-direction. The magnetic field was also set to be uniform with a value of \SI{2.5}{\tesla}. The neutral tritium background gas was set to be uniformly distributed with a value of \SI{1e22}{\per\cubic\m} A total of 1600 events were used per simulation.}

\begin{figure}[tbp]
  \centering
  \includegraphics[width=0.8\textwidth]{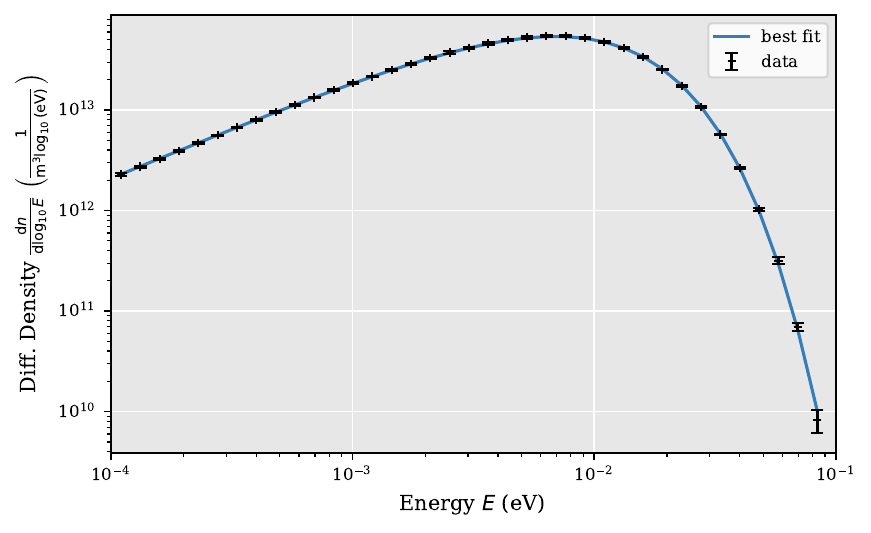}
  \caption{Electron spectrum and fit. Monoenergetic electrons are injected with energies of $E_\text{inj.} = \SI{8}{meV}$ into a neutral gas reservoir with a Maxwell-Boltzmann distributed velocity $(T = \SI{80}{K}$). No other interactions are activated than elastic scattering and the cross section of elastic scattering is set to a constant value. The data from the numerical experiment is fitted by a scaled Maxwell-Boltzmann distribution ($T_\text{fit} = \SI{80.40 \pm 0.08}{K}$). The error bars are calculated by using six simulations with different random seeds \changed{but otherwise unchanged parameters}. The $R^2$ value for the fit is 0.9998.}
  \label{fig:boltzmann_test}
\end{figure}

If elastic scattering is implemented correctly, then these particles will adopt the velocity distribution of the neutral gas independent of the initial energy. The shape can be described by the Maxwell-Boltzmann distribution \citep{peckham.1992}, scaled by the factor $\sqrt{E}$ taking into account the logarithmic binning and the speed of particles leaving the simulation domain. The simulation result and a fit to the data are shown in in Figure~\ref{fig:boltzmann_test}. 
It can be seen that the data is well described by the fit with little deviations over a wide range of energies. The deviations can be attributed to the inherent statistical uncertainty of the Monte Carlo method. Simulations with other injection energies resulted in the same distribution (not shown). 
%It can therefore be concluded that elastic scattering, and therefore the general structure of the code is implemented correctly.

\begin{figure}
\centering
\includegraphics[width=1\textwidth]{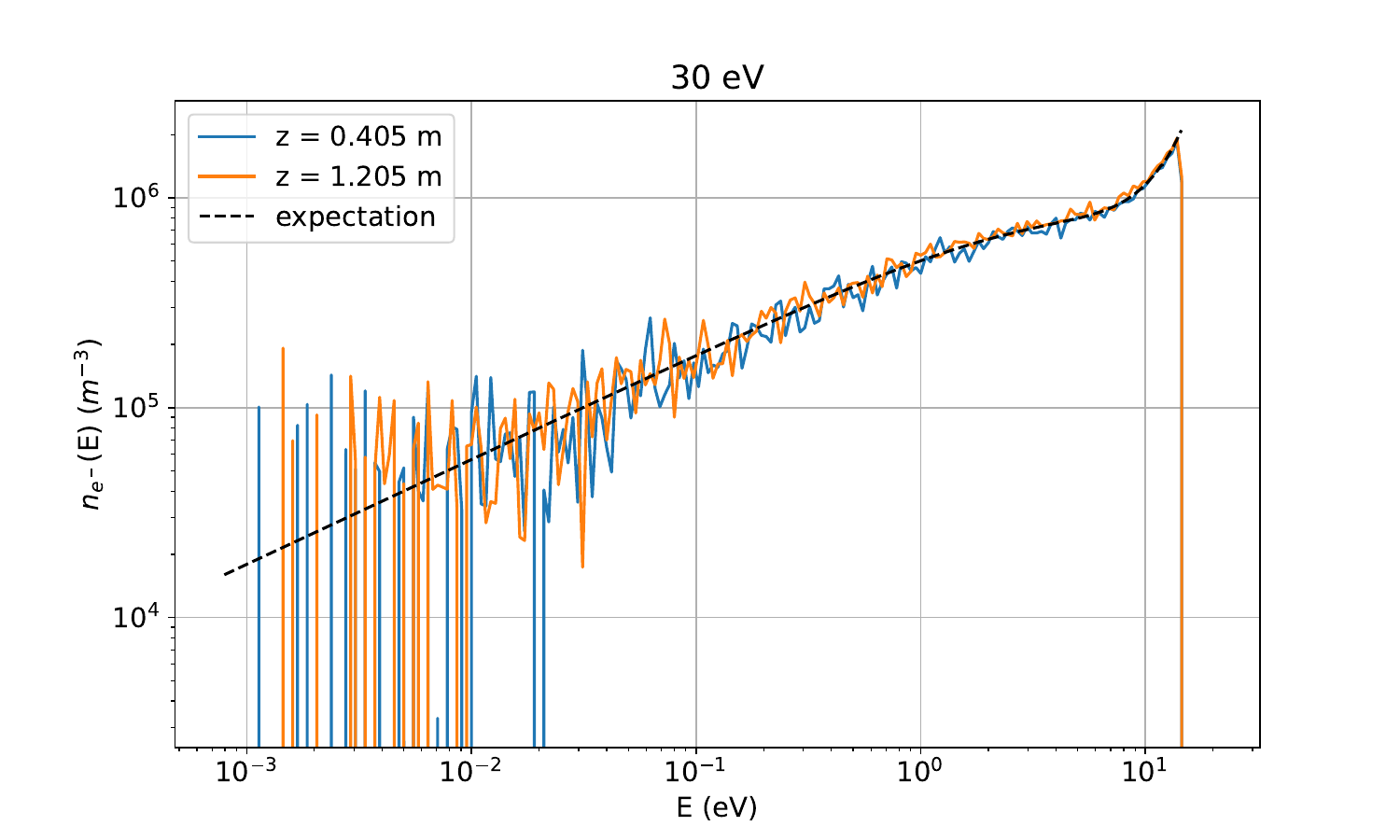}
\caption{Energy spectrum for electrons being injected in the middle of the simulation domain at positions symmetrically placed around the injection site. The electrons were injected with an energy of \SI{30}{eV}. The black dashed curve shows the expectation for the spectrum.}
\label{fig:ionization_test}
\end{figure}

\paragraph{Ionization}
\changed{One way to test the validity of the ionization process is by injecting monoenergetic particles and taking a look at the energy distribution after they interact with the background gas. In case of a high injection energy there is a high likelihood for multiple ionization processes to occur, which would complicate the prediction of the shape of the energy spectrum. In order to guarantee only a single ionization event per particle can take place the energy has to be chosen below twice the ionization threshold. Since the energy distribution of one of the electrons after the ionization process is known it is possible to calculate the expected energy distribution of the electrons. However, due to the different angle distributions the population of scattered and ejected electrons have different velocities in $z$-direction and thus influence the density in a different way, which complicates the calculation for the expectation. In order to remedy this problem it was decided that no magnetic field will be used in the simulation, which simply lets electrons with a low $z$-velocity escape in radial direction.\\
For the test an energy of \SI{30}{eV} has been chosen, since it is close to the maximum value of the cross section as well as only allowing for a single ionization event to take place. Electrons are injected in the middle of the simulation domain with an isotropic velocity direction. The density of the target gas was set to \SI{2.7e21}{\per\cubic\m} with an uniform distribution across the whole domain. In total a number of \num{1.2e6} primary electrons were simulated.\\
Figure~\ref{fig:ionization_test} shows the electron spectrum at two different $z$-positions some distance away from the injection site as well as the expectation for the shape of the electron energy spectrum. It can be seen, that while the expectation is met, the results are quite noisy. Simulations in which the angle distributions excluded scatterings to low $v_z$ values showed much smoother results. These results are not shown, since they require manually changing the scattering angle distribution in the source code and can thus not be reproduced as easily.
}

\begin{figure}
\centering
\includegraphics[width=1\textwidth]{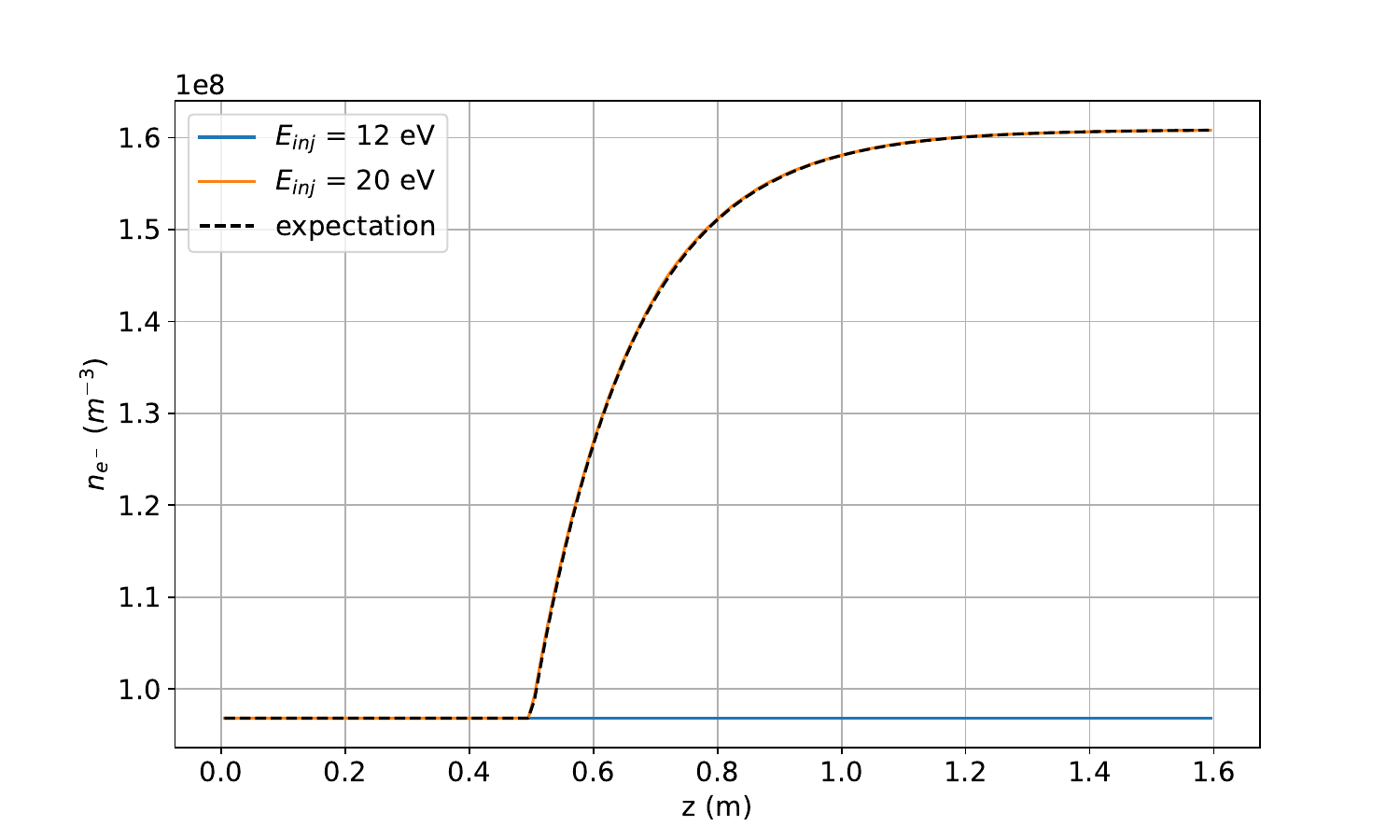}
\caption{Electron densities for two injection energies, one below the threshold of \SI{12.754}{eV} for the interaction and one above. The densities were scaled to be of the same magnitude for $z < 0.5$.}
\label{fig:excitation_test_densities}
\end{figure}

\begin{figure}
\centering
\includegraphics[width=1\textwidth]{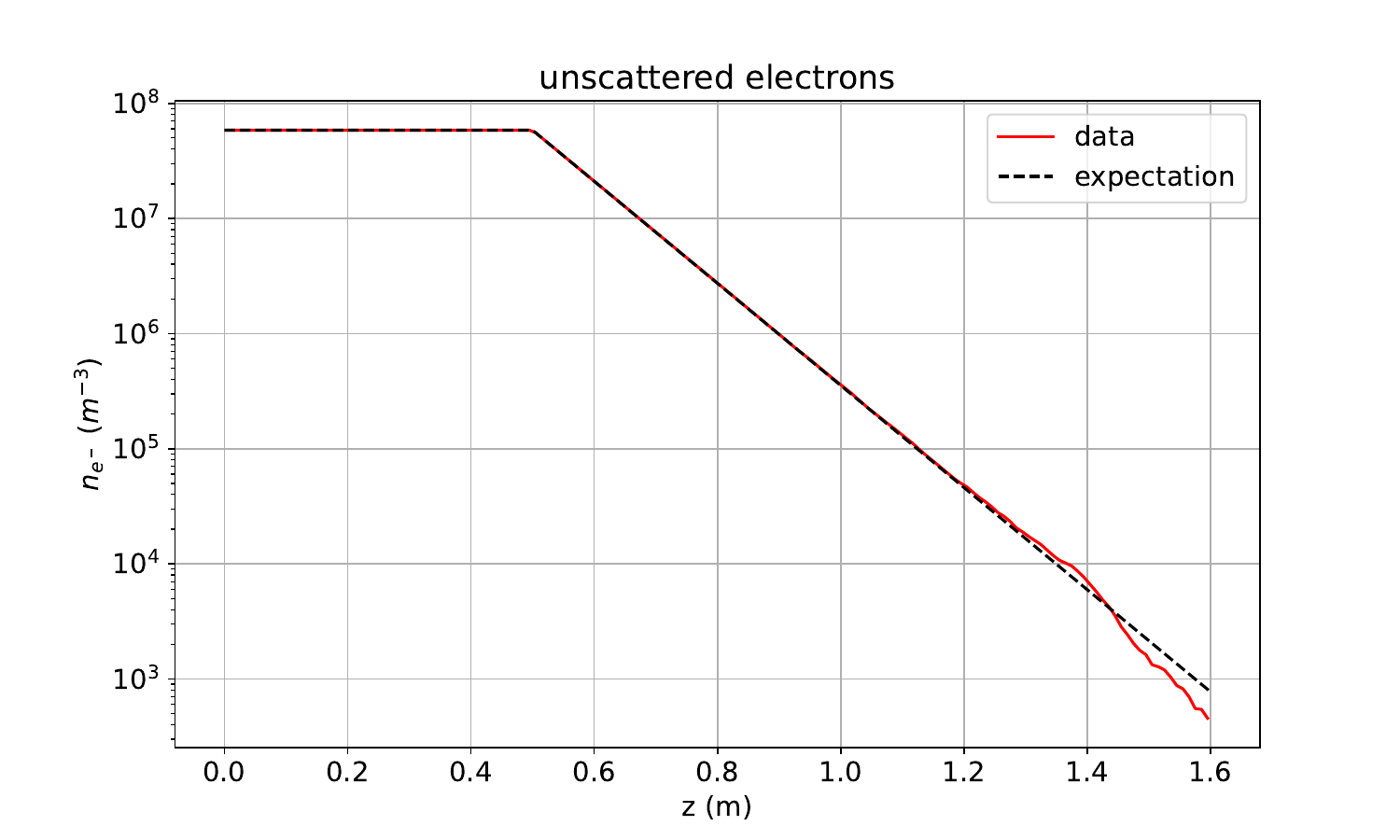}
\caption{Density of unscattered electrons for $E_\text{inj} = \SI{33}{eV}$.}
\label{fig:excitation_double_scattering_unscattered}
\end{figure}

\begin{figure}
\centering
\includegraphics[width=1\textwidth]{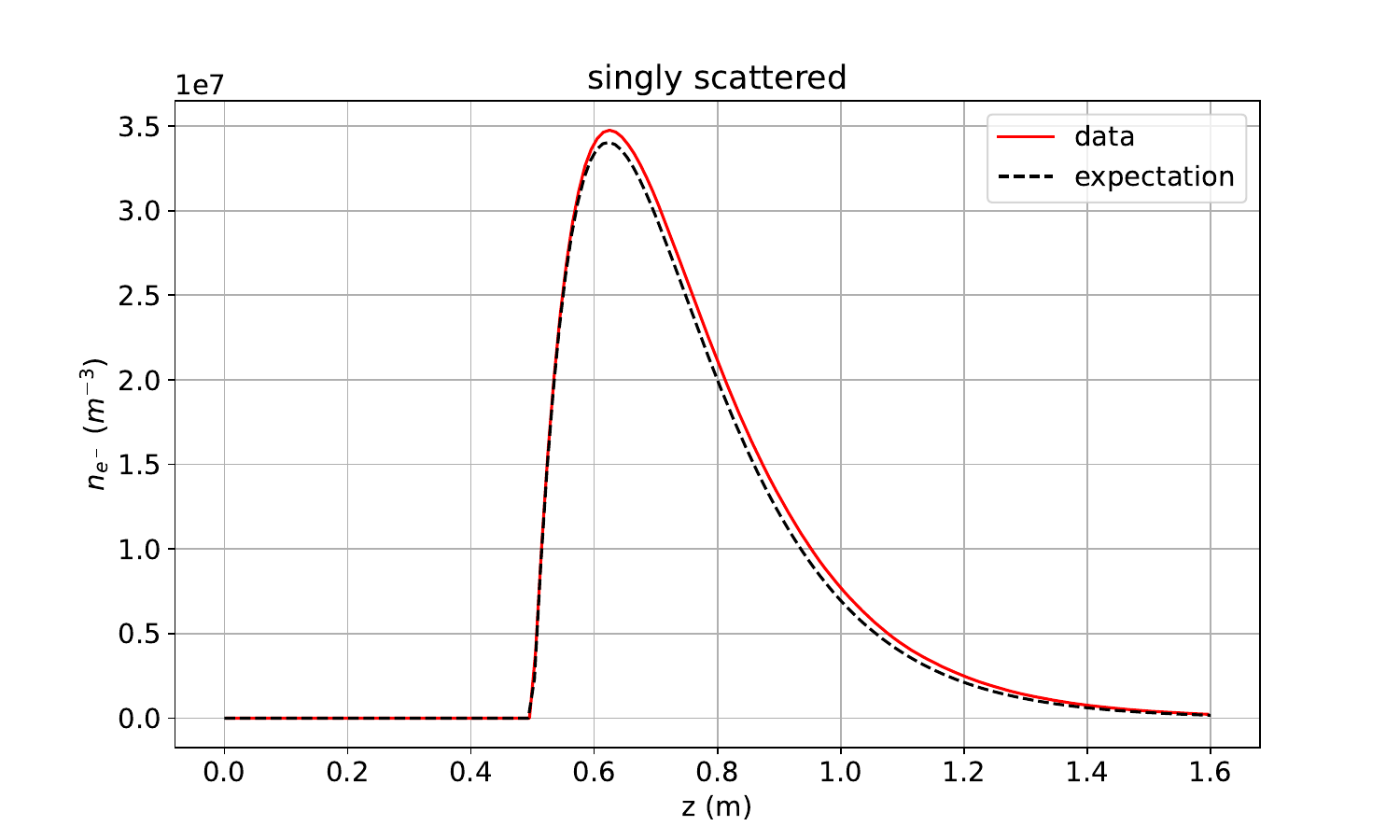}
\caption{Density of singly scattered electrons for $E_\text{inj} = \SI{33}{eV}$.}
\label{fig:excitation_double_scattering_singly}
\end{figure}

\begin{figure}
\centering
\includegraphics[width=1\textwidth]{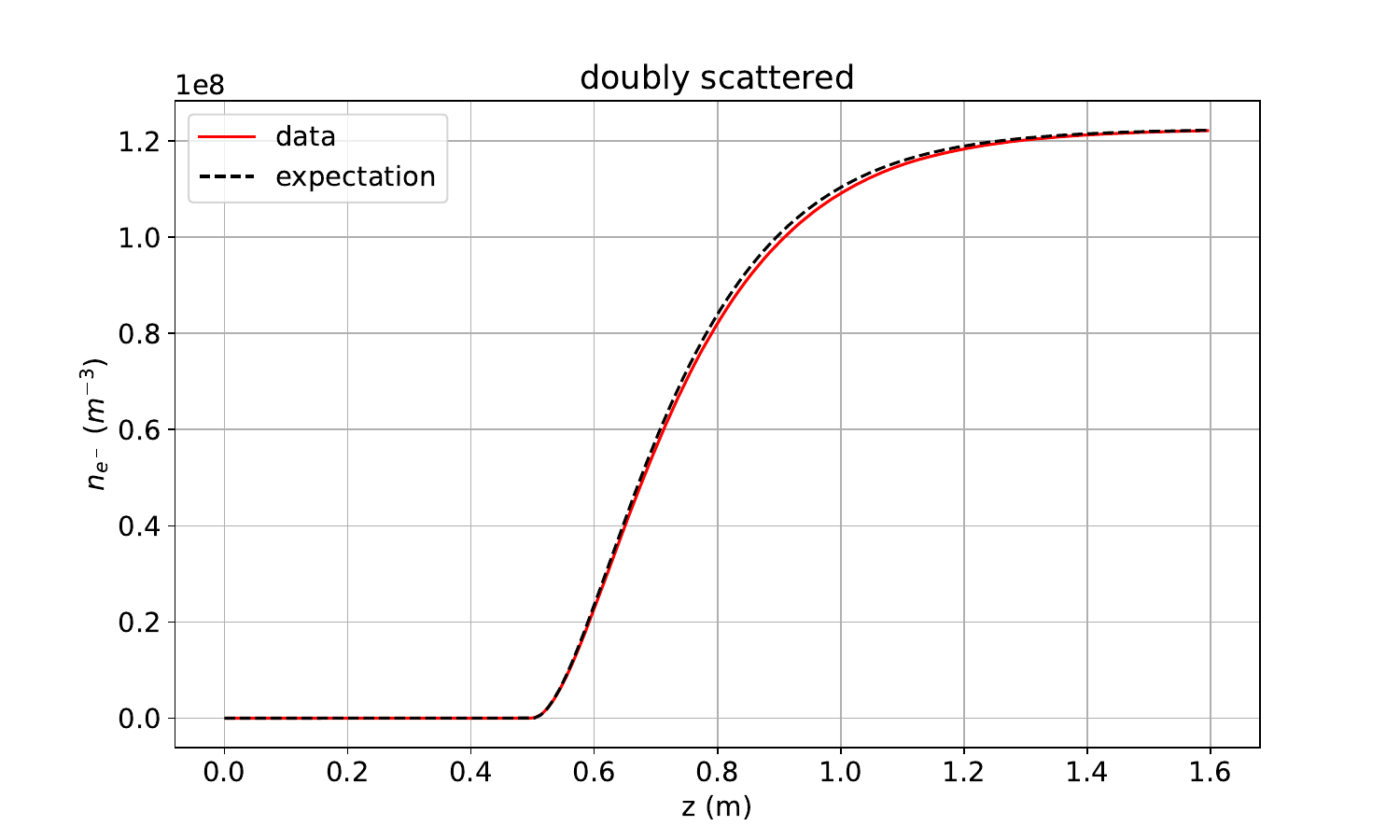}
\caption{Density of doubly scattered electrons for $E_\text{inj} = \SI{33}{eV}$.}
\label{fig:excitation_doubly_scattering_doubly}
\end{figure}

\begin{figure}
\centering
\includegraphics[width=1\textwidth]{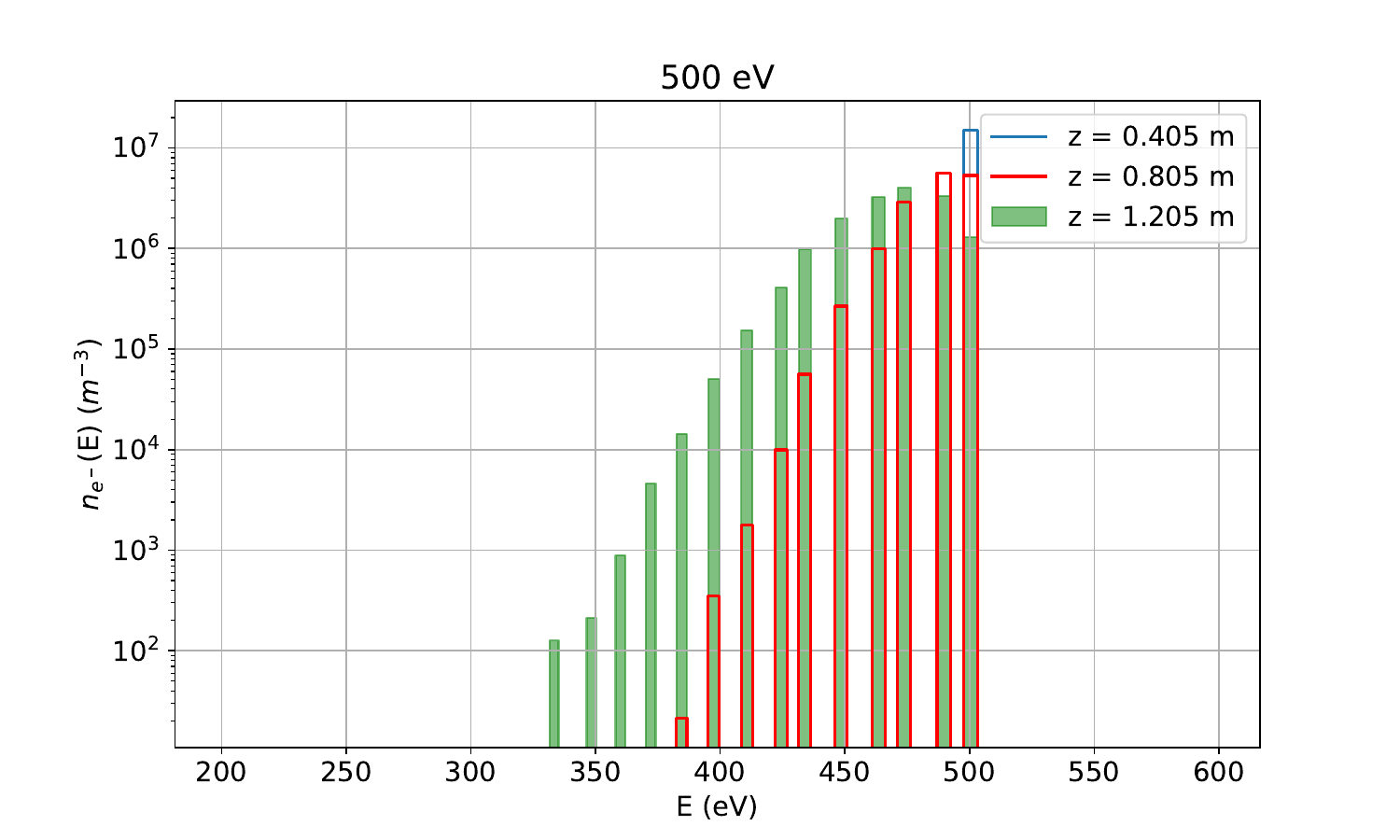}
\caption{Electron energy spectrum for $E_\text{inj} = \SI{500}{eV}$ at three different $z$-positions. The neutral density is non-zero only for $z>\SI{0.5}{\m}$.}
\label{fig:excitation_spectrum_500eV}
\end{figure}

\paragraph{Excitation}
\changed{The excitation process is somewhat similar to the ionization process, in that it requires a threshold energy of the impact particle. The main difference is that no new particles are created. Due to the assumption made on the scattering angle distribution in the excitation process (see section~\ref{ssec:interactions}), it is quite simple to formulate an analytic example if only a single interaction is possible during propagation.\\
There is a high amount of potential excitation interactions that electrons can have with molecular tritium. However, the only difference between these interactions are the energy threshold $E_\text{th}$ and the cross section of the interaction. Therefore it is sufficient to only consider a single implemented excitation process in order to test the validity of the approach.\\
In the setup of the simulations, the electrons were injected at the left side of the simulation domain with a specific energy and their velocity aligned with the $z$-axis. The neutral density $n_\text{n}$ is chosen to be uniformly distributed within the source starting from a position $z_\text{neutral} > 0$ in order to allow for a distance of free propagation. The value of $n_\text{n}$ is chosen to be \SI{3.21e21}{\per\cubic\m} in order to allow for an appropriate MFP. The used interaction (or rather the specific cross section) has a threshold energy $E_\text{th} = \SI{12.754}{eV}$. The injection energies considered were \SI{12}{eV}, \SI{20}{eV}, \SI{33}{eV} and \SI{500}{eV}. The first energy was chosen such as to test for the threshold energy, the second one allows for a singular interaction, the third for at most two interactions and the fourth one allows for many interactions. For each case a number of \num{8e5} particles are simulated.\\
The density of the injected particles $n_u$ is expected to follow an exponential distribution, depending on the mean free path of the interaction
\begin{equation}
\label{eq:density_injected}
n_u = n_0 \exp\left(-\frac{z - z_\text{neutral}}{\lambda_\text{mfp}(E_\text{inj})}\right) \commaeq
\end{equation}
with $n_0$ being the density in case no interaction would have been performed.
This is valid for any injection energy $E_\text{inj}$.\\
In case only a single scattering is possible the density of the scattered particles $n_s$ is given by
\begin{equation}
n_s \propto \left(1-\exp\left(-\frac{z - z_\text{neutral}}{\lambda_\text{mfp}(E_\text{inj})}\right)\right) n_0 \fullstop
\end{equation}
Due to the predictable change in velocity as well as the conserved direction of motion it is possible to predict the proportionality factor from equation~\ref{eq:target_field_density}. This leads to 
\begin{equation}
\label{eq:density_single_scattering}
n_s = \left(1-\exp\left(-\frac{z - z_\text{neutral}}{\lambda_\text{mfp}(E_\text{inj})}\right)\right) \sqrt{\frac{E_\text{inj}}{E_\text{inj} - E_\text{th}}}n_0 \fullstop
\end{equation}
Adding $n_u$ and $n_s$ yields the expected density profile.\\
In case the impact energy allows for at most two excitation events to be performed, the density for the twice scattered electrons $n_{s,2}$ can be calculated as well. In order to do so, one just has to calculate the probability $p(z)$ for a single particle to scatter twice within the length $z$:
\begin{equation}
p(z) = \int_0^z dx \int_0^{z-x} dy\, p_1(x) p_2(y) \commaeq
\end{equation}
where $p_1(x)$ is the probability for a particle to scatter after a distance $x$ and $p_2(y)$ is the same at the new energy after the first scattering. Since the distribution of distances follows an exponential distribution, this leads to
\begin{equation}
p(z) = \int_0^zdx \int_0^{z-x}dy\, \left[\frac{1}{\lambda_1} \exp\left(-\frac{x}{\lambda_1}\right)\right] \left[\frac{1}{\lambda_2} \exp\left(-\frac{y}{\lambda_2}\right)\right] \commaeq
\end{equation}
where $\lambda_1$ is the mean free path for the injected electron energy $E_{inj}$ and $\lambda_2$ for $E_{inj} - E_{th}$. Together with the requirement that a particle has to either scatter once, twice or not at all leads to
\begin{equation}
\begin{split}
n_s &= n_0\sqrt{\frac{E}{E - E_{th}}} \frac{\lambda_2}{\lambda_2 - \lambda_1} \left( \exp\left(-\frac{z}{\lambda_2}\right) - \exp\left(-\frac{z}{\lambda_1}\right)\right) \\
n_{s,2} &= n_0 \sqrt{\frac{E}{E - 2E_{th}}} \left[1-\left(1 - \frac{\lambda_2}{\lambda_2 - \lambda_1}\right) \exp\left(-\frac{z}{\lambda_1}\right) - \frac{\lambda_2}{\lambda_2 - \lambda_1}\exp\left(-\frac{z}{\lambda_2}\right)\right] \fullstop
\end{split}
\end{equation}
For the case that both $\lambda_1$ and $\lambda_2$ have the same value the expressions change but can be derived the same way.
\\
For higher interaction counts this becomes increasingly more difficult to formulate an expectation. Therefore only the case for a single scattering are shown in Figure~\ref{fig:excitation_test_densities}. In this plot the \SI{12}{eV} is also shown to indicate the value of $n_0$. Furthermore, the density for the \SI{20}{eV} case has been scaled such that it fits to $n_0$ in the non-interaction region. It can be seen that the data fits the expectation very well, with the relative error being below 0.5\% everywhere. In Figures~\ref{fig:excitation_double_scattering_unscattered},
\ref{fig:excitation_double_scattering_singly} and \ref{fig:excitation_doubly_scattering_doubly} the densities for the electrons that have scattered a zero, once and twice are shown respectively for the case of a maximum of two interactions. While the unscattered electrons follow the expectation quite well, the singly and doubly scattered electrons deviate somewhat from the expectation. This might be due to differences in the calculated cross section with the one calculated in the code. In Figure~\ref{fig:excitation_spectrum_500eV} the energy spectrum for an injection energy of $E_\text{inj} = \SI{500}{eV}$ is shown. Overall, it can be seen that the length of the simulation domain is not large enough for any electron to scatter down to below the threshold energy. However a shift in the energy with the maximal density can be seen, as would be expected.
}

\begin{figure}
\centering
\includegraphics[width=1\textwidth]{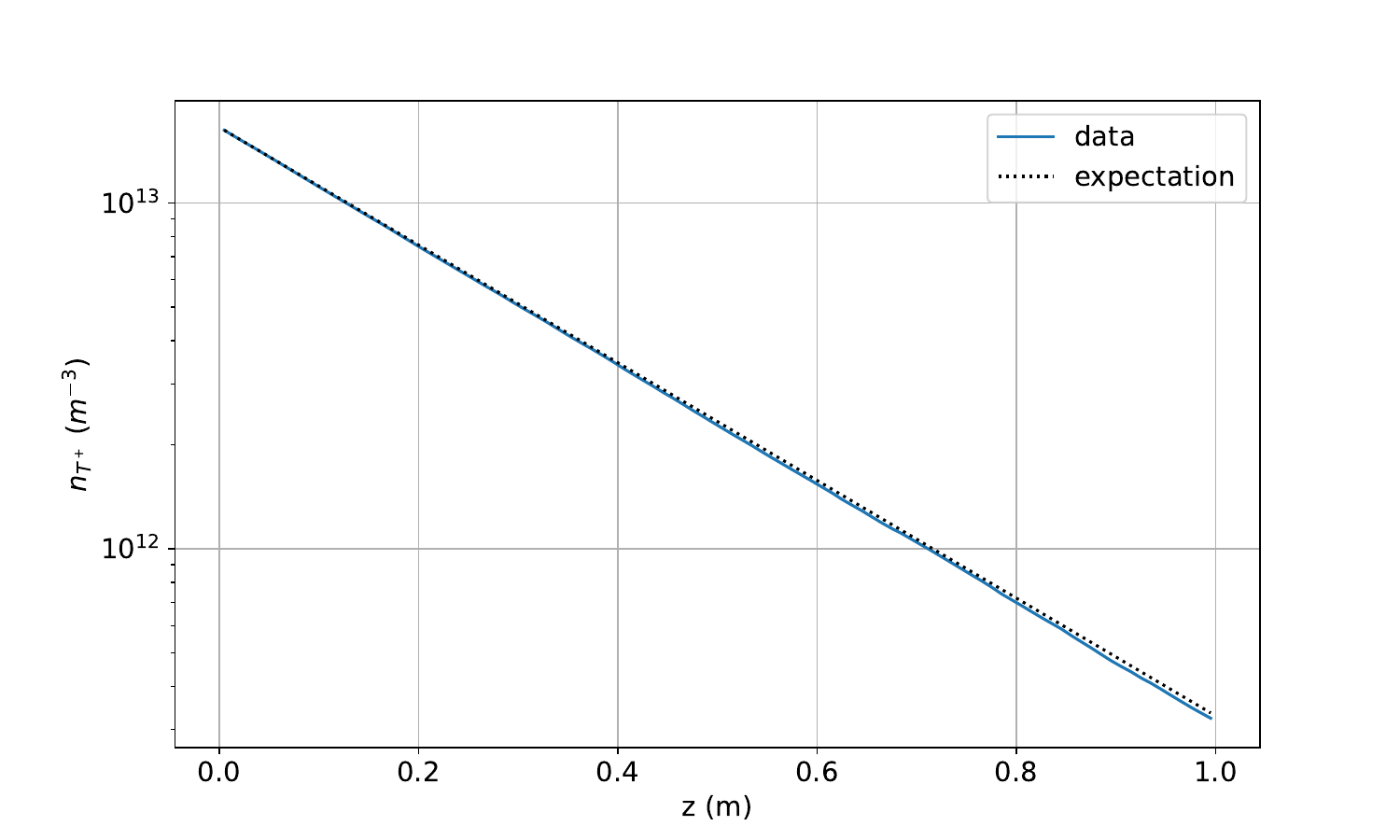}
\caption{Tertiary Clustering of T$^+$ with T$_2$. Neutral gas density was set to \SI{1e22}{\per\cubic\metre}. Ions were injected at $z = 0$ with an energy of \SI{6.8}{\milli eV} in the direction of the $z$-axis.}
\label{fig:clustering_tp_ter}
\end{figure}

\begin{figure}
\centering
\includegraphics[width=1\textwidth]{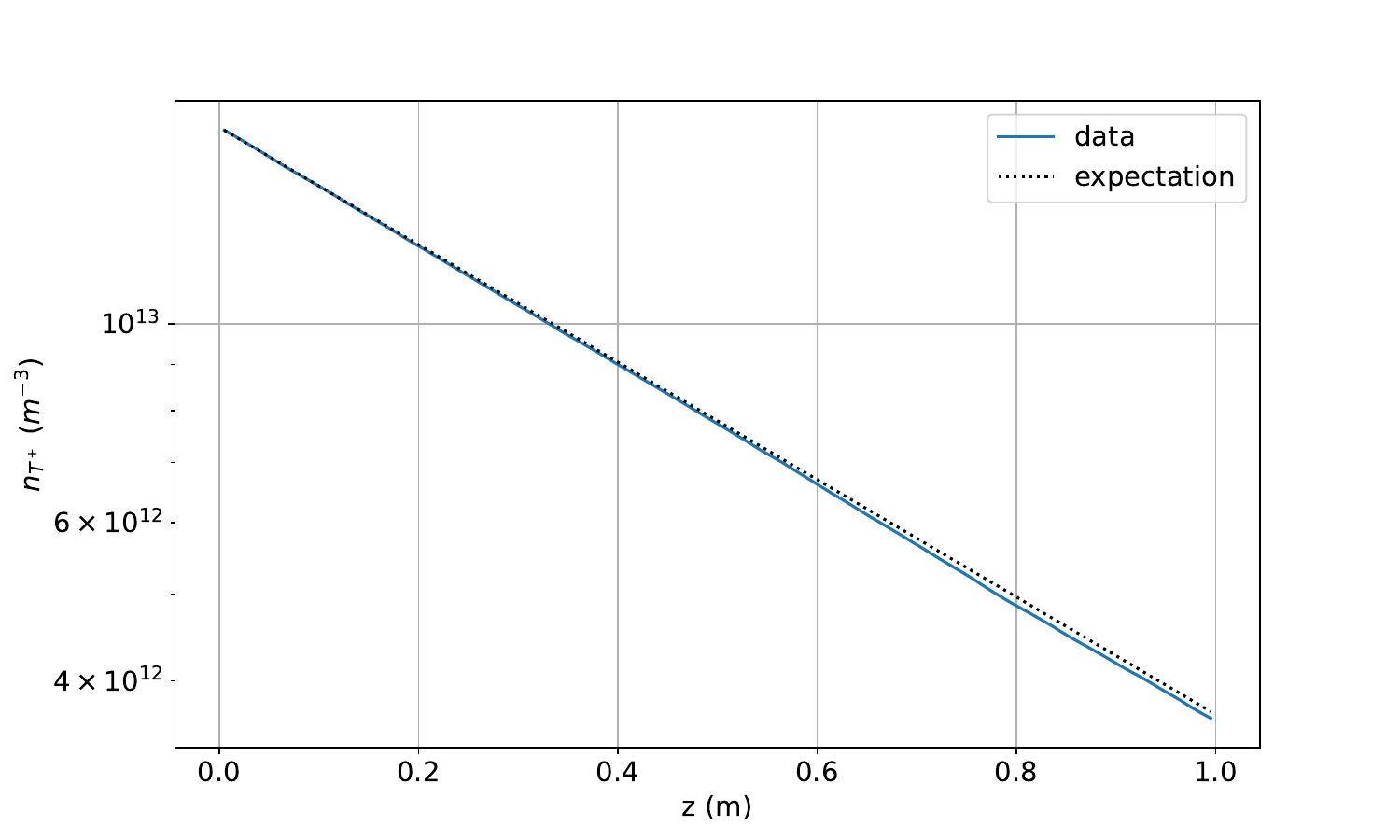}
\caption{Radiative Clustering of T$^+$ with T$_2$. Neutral gas density was set to \SI{2e25}{\per\cubic\metre}. Ions were injected at $z = 0$ with an energy of \SI{6.8}{\milli eV} in the direction of the $z$-axis.}
\label{fig:clustering_tp_rad}
\end{figure}

\begin{figure}
\centering
\includegraphics[width=1\textwidth]{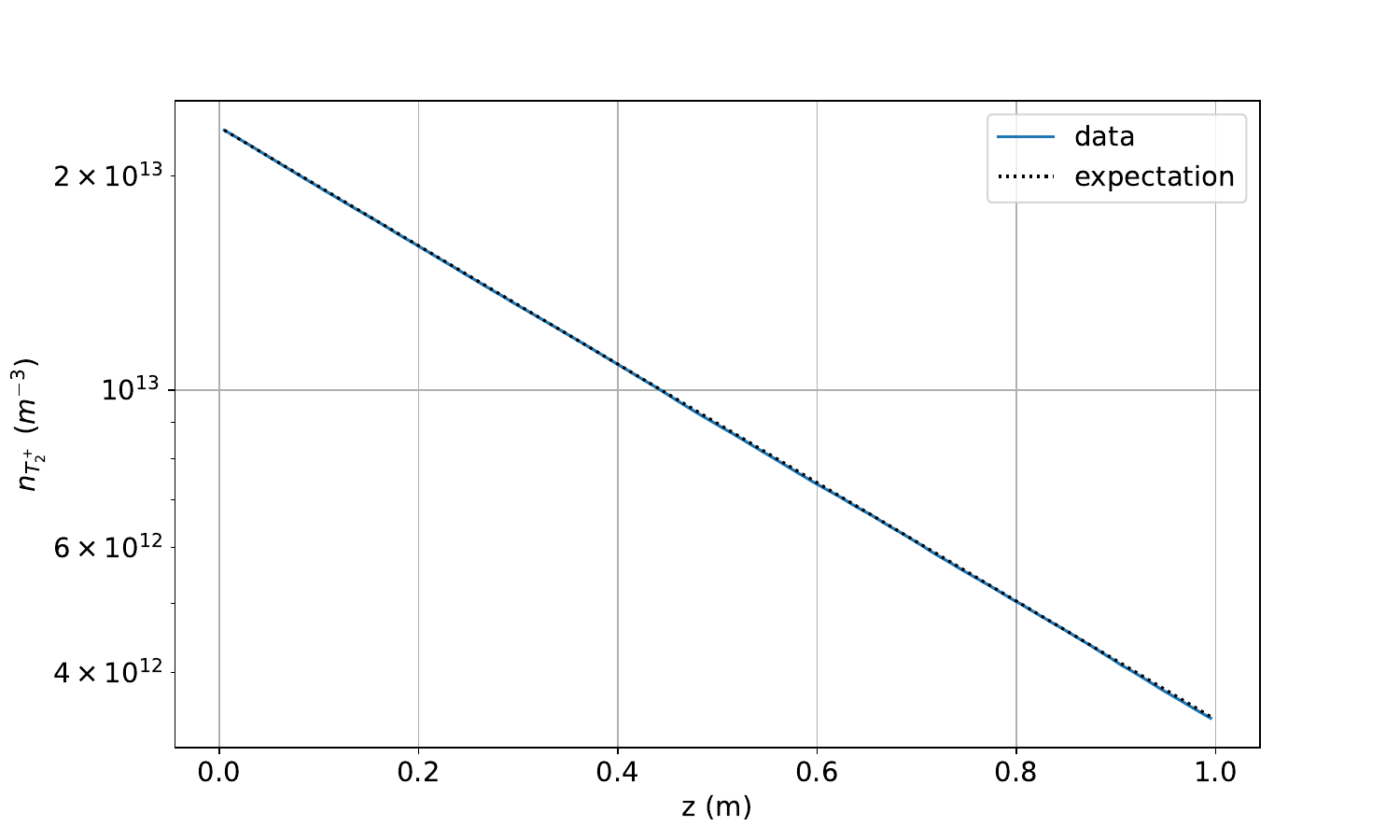}
\caption{Radiative Clustering of T$_2^+$ with T$_2$. Neutral gas density was set to \SI{1e18}{\per\cubic\metre}. Ions were injected at $z = 0$ with an energy of \SI{6.8}{\milli eV} in the direction of the $z$-axis.}
\label{fig:clustering_t2p}
\end{figure}

\paragraph{Clustering}
\changed{In the case of an ion-beam moving with a constant direction and velocity through an background gas, with which it can interact via clustering, it is possible to predict the form of the density of the ion beam
\begin{equation}
n_\text{ion} \propto \exp\left(-\frac{z}{\lambda_{\text{mfp}}}\right) \fullstop
\end{equation} 
Since both the profile of the background gas density, the energy of the ions and the cross section of the interactions are known it is easily possible to check if the simulation corresponds to the analytical values.\\
For this test case only the interacting particles are considered, since during the interaction the velocity direction of the products becomes essentially randomized, which makes a prediction more challenging.\\
As described in section~\ref{ssec:interactions}, there are two types of clustering: radiative and tertiary. In case of the tertiary process only rate coefficients are known, s.t. the cross section is only approximated. This leads to 
\begin{equation}
\label{eq:clustering_test_tertiary_mfp}
\lambda_{\text{mfp, ter}} = \frac{v}{k_\text{ter} n_{\text{T}_2}^2}
\end{equation}
for the mean free path, where $v$ is the velocity of the ion in the rest-frame of the tritium gas, $k_\text{ter}$ is the rate tertiary rate coefficient and $n_{\text{T}_2}$ is the density of the background gas.\\
For the radiative clustering there are two types of cross sections, first for T$^+$ only the rate coefficient is known again, while for T$_2^+$ an expression for the actual cross section is known. This means that there are two different expressions for the mean free path:
\begin{equation}
\label{eq:clustering_test_radiative_rateCoefficient}
\lambda_{\text{mfp,rad}} = \frac{v}{k_\text{rad} n_{\text{T}_2}}
\end{equation}
and
\begin{equation}
\label{eq:clustering_test_radiative_cross}
\lambda_{\text{mfp,rad}} = \frac{1}{\sigma n_{\text{T}_2}} \fullstop
\end{equation}
All three cases were tested using the same setup. The ions were injected on the left side of the simulation domain with an energy of \SI{6.8}{\milli eV} and a direction in positive $z$-direction. The neutral gas was chosen to be uniformly distributed in the simulation domain. A number of \num{8e4} events were chosen to be simulated. Furthermore a strong magnetic field of \SI{2.5}{\tesla} was applied. The difference between the simulations is solely the density of the neutral gas due to the different magnitudes of the MFP. Therefore a density of \SI{1e18}{\per\cubic\m} has been chosen for the reaction T$_2^+$ + T$_2$. The cross section for this reaction has a value of $\sigma = \SI{1.91e-18}{\square\m}$. For the tertiary process a density of \SI{1e22}{\per\cubic\m} was chosen. The value of the rate coefficient at \SI{80}{\kelvin} is $ k_\text{ter} = \SI{2.6e-41}{\m\tothe{6}\per\second}$. And lastly for the radiative process of T$^+$ a density of \SI{2e25}{\per\cubic\m} has been chosen with a radiative association rate coefficient of $k_\text{rad} = \SI{5e-23}{\cubic\m\per\s}$ at \SI{80}{\kelvin}.\\
The results of the tree simulations together with the expected result are shown in Figures~\ref{fig:clustering_tp_ter}, \ref{fig:clustering_tp_rad} and \ref{fig:clustering_t2p}.\\
As can be seen the expectation and the results fit quite well, only showing some deviation in the low density region due to statistical noise. The values for the rate coefficients and cross section were also inferred and have shown a deviation below 1\% for all three cases.}

\begin{figure}
\centering
\includegraphics[width=1\textwidth]{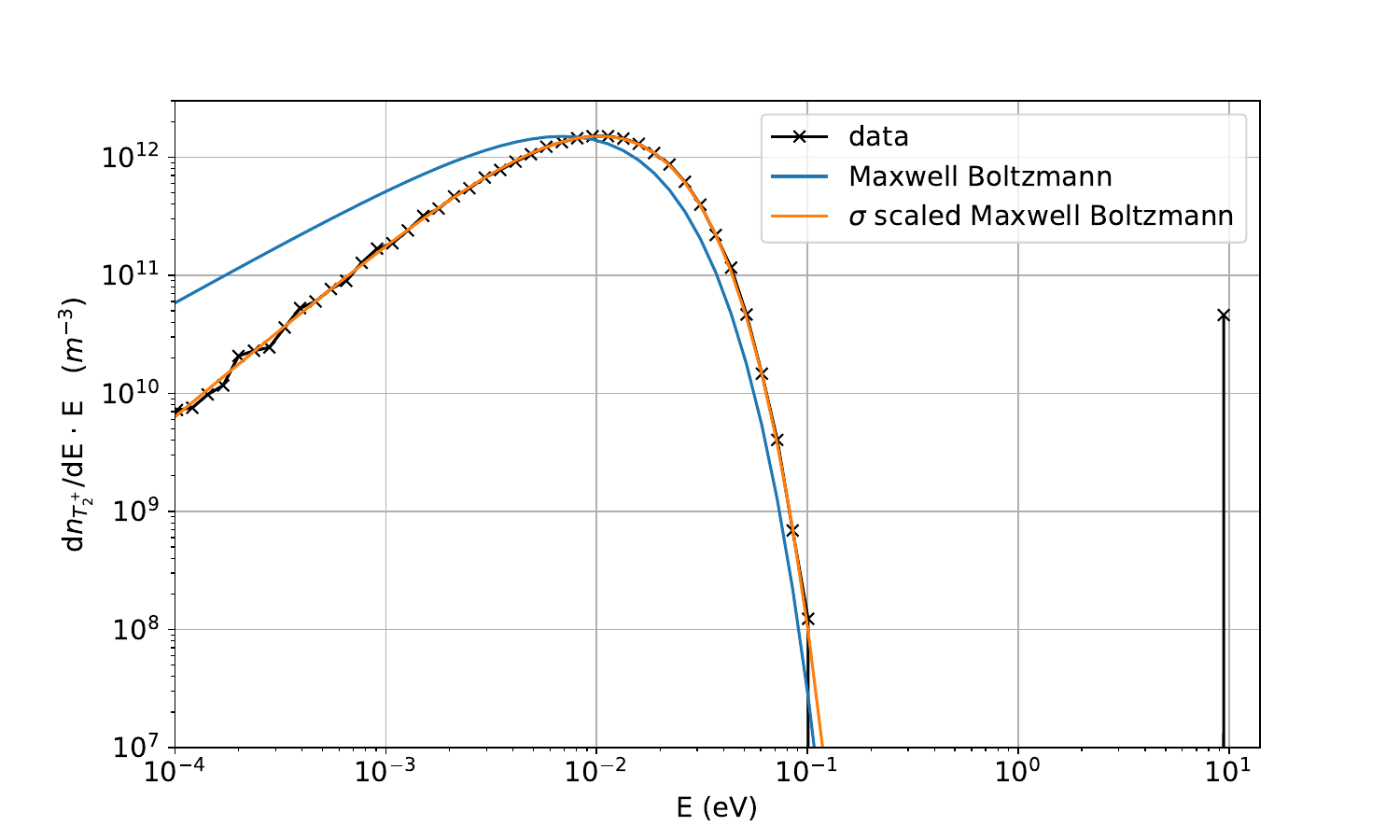}
\caption{Energy spectrum of T$_2^+$ with the only interaction being charge exchange with the T$_2$ background gas. The Maxwell Boltzmann distribution is not met, but scaled with $1/\sigma(E)$ yields a near perfect fit. This is due to a higher likelihood of scattering for lower energetic particles due to the cross section. A population of unscattered ions can be seen at \SI{10}{eV}.}
\label{fig:charge_exchange}
\end{figure}

\paragraph{Charge Exchange}
\changed{The expectation for charge exchange is quite similar as for the case of elastic scattering, namely that the initial population adopts the energy distribution of the background population. The difference being that only one scattering event per particle is needed to thermalize.\\
For the simulation monoenergetic T$_2^+$ ions are injected from the left side of the simulation domain with an energy of \SI{10}{eV} and a velocity direction pointing in positive $z$-direction. The neutral density has been set to \SI{1.224e19}{\per\cubic\m} to allow for a short MFP with a temperature of \SI{80}{\kelvin}.\\
The energy spectrum for the ions is shown in Figure~\ref{fig:charge_exchange} close to the right side of the simulation domain at $z = \SI{1.44}{\m}$. A peak at \SI{10}{eV} can also be seen, due to unscattered particles. Furthermore, the Maxwell Boltzmann distribution is not well met. This is however an effect of a non-constant cross section. From the cross section shown in Figure~\ref{fig:t2p_interactions} it becomes apparent that lower energetic ions have a higher likelihood to interact and therefore spending a lower amount of time at their current energy than a particle with a higher energy before scattering. Assuming the time to scale inversely with the probability of performing an interaction leads to an additional factor in the Maxwell Boltzmann distribution of $1/\sigma(E)$. This scaled version is also shown in the plot of the results and it is evident that this expectation is met quite well confirming the above discussion.
}

\begin{figure}
\centering
\includegraphics[width=1\textwidth]{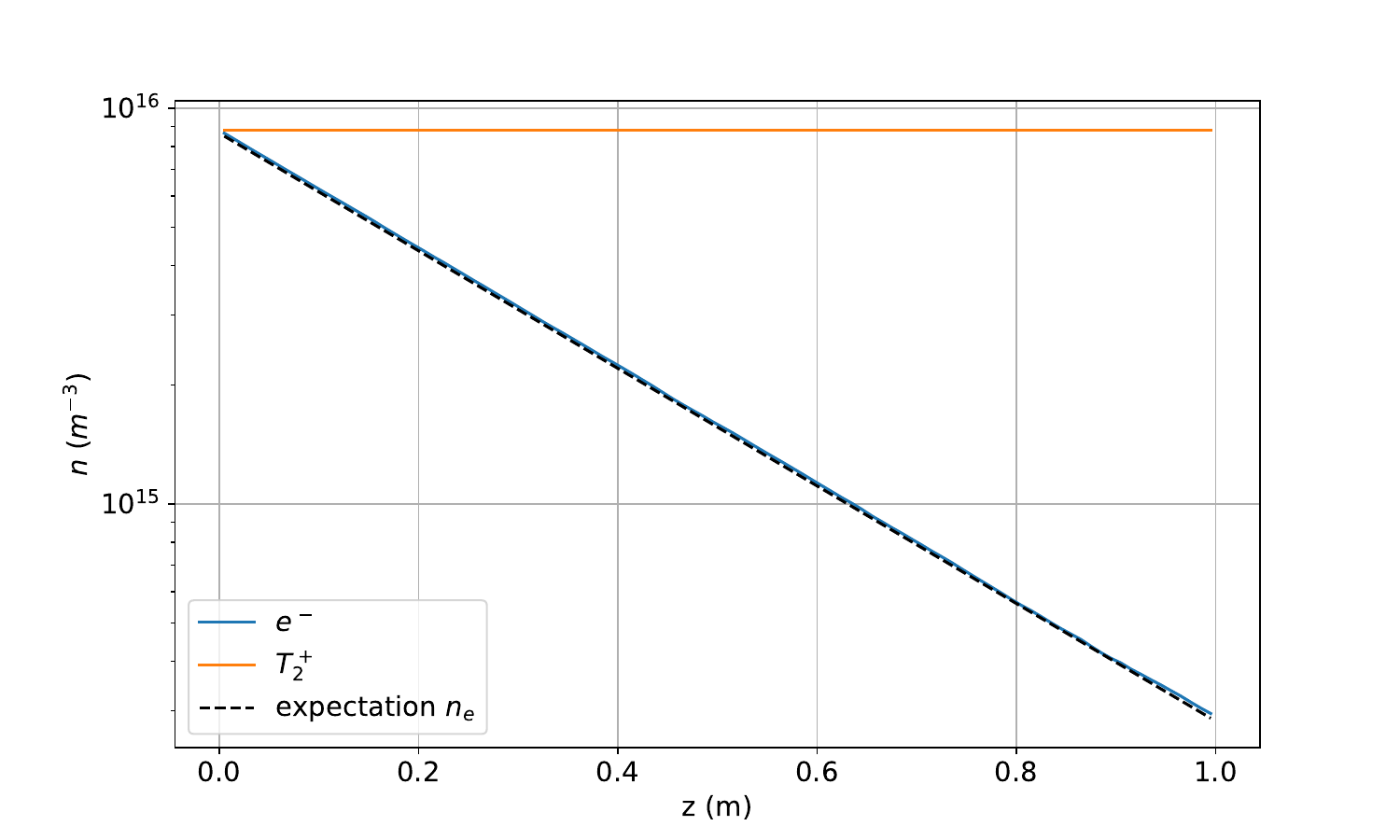}
\caption{Particle densities for recombination of e$^-$ with T$_2^+$. Only electrons experience recombination, while the ions act as a constant background.}
\label{fig:recombination_e_only}
\end{figure}

\begin{figure}
\centering
\includegraphics[width=1\textwidth]{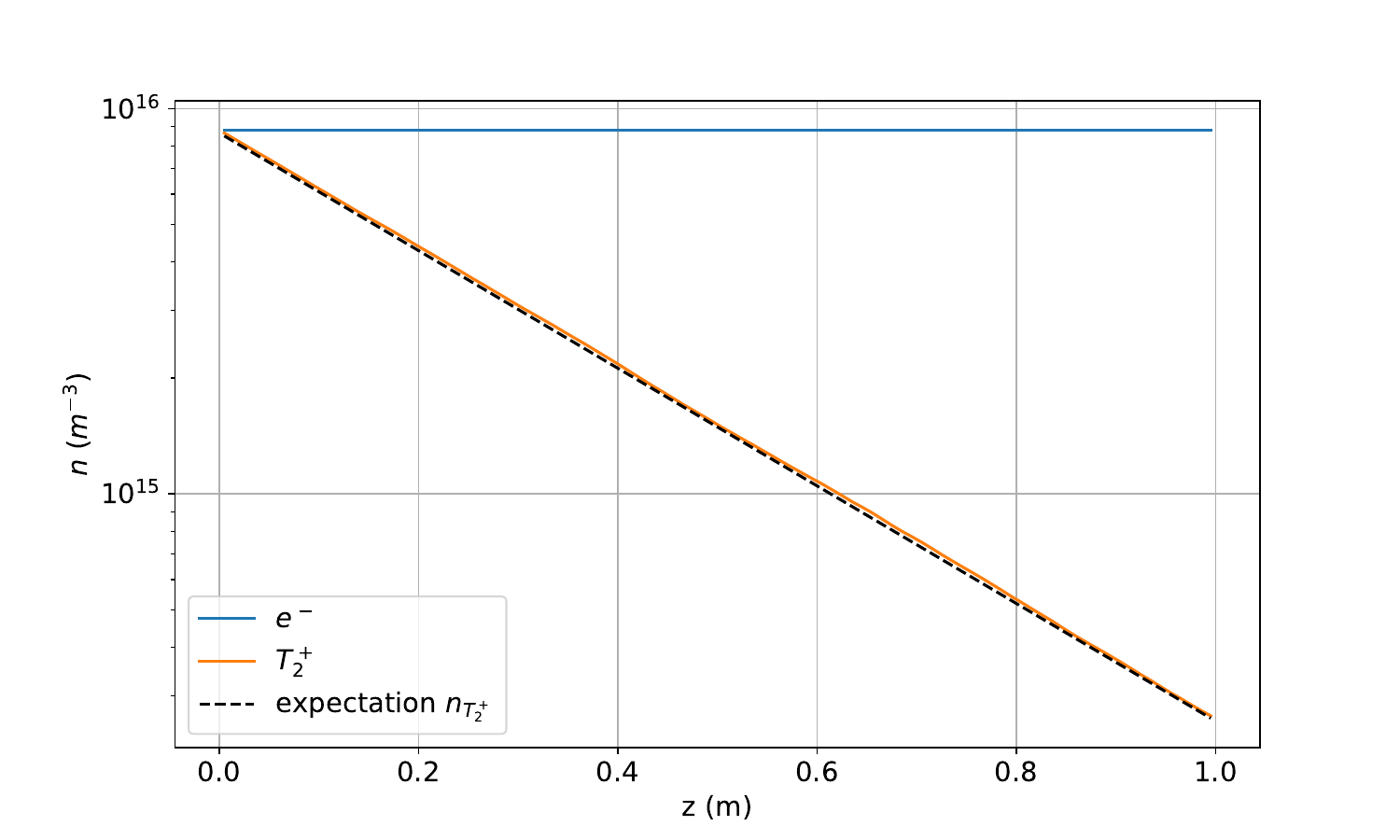}
\caption{Particle densities for recombination of e$^-$ with T$_2^+$. Only ions experience recombination, while the electrons act as a constant background.}
\label{fig:recombination_t2p_only}
\end{figure}

\begin{figure}
\centering
\includegraphics[width=1\textwidth]{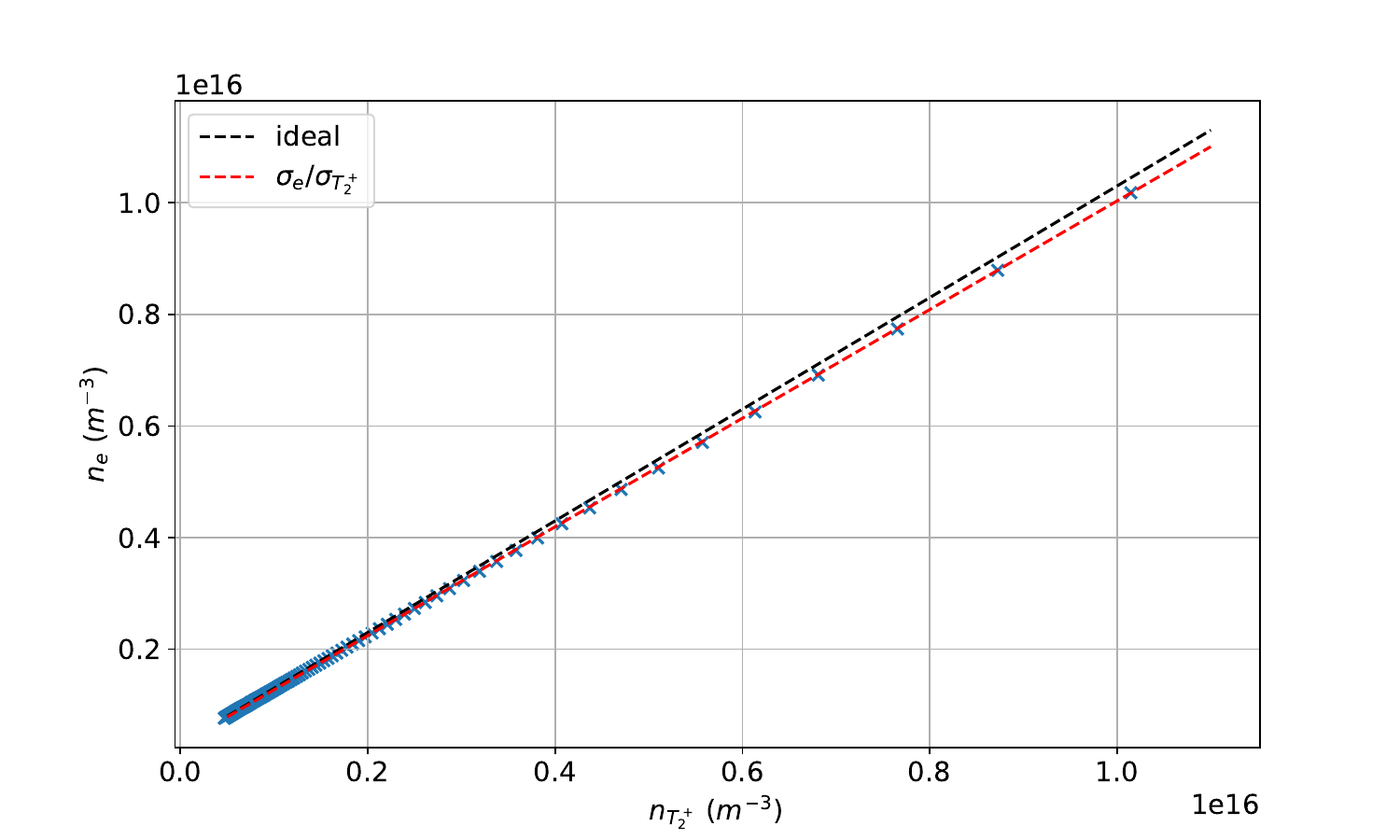}
\caption{Particle densities for recombination of e$^-$ with T$_2^+$. Dependence between the two particle densities follow the expected linear relationship. The slope of the red line is determined from the way the cross sections are calculated within the code with $\sigma_e/\sigma_{\text{T}_2^+} = 0.973$.}
\label{fig:recombination_ne_nt2p}
\end{figure}

\paragraph{Recombination}
\changed{In order to test recombination there are two parts that were tested separately. First recombination was tested for both interaction partners against a fixed background density to retrieve the cross section and check if the simple expected exponential profile is retrieved correctly. Second, recombination with both interaction partners being fully variable was simulated. For this case it is not as simple to formulate an expectation for the density profiles in relation to the spatial coordinates. However, it is possible to set the densities of both interaction partners into relation. Therefore one starts with a simple advection equation with a sink term that couples the two interaction partners
\begin{equation}
\begin{split}
-\partial_t n_e &= v_e\partial_z n_e + \sigma(E_e)n_en_iv_e\\
-\partial_t n_i &=v_i\partial_z n_i + \sigma(E_i)n_en_iv_e \fullstop
\end{split}
\end{equation}
Since only the equilibrium is of interest, the time derivative will vanish. This leads to 
\begin{equation}
\label{eq:ne_nt2p}
n_e = \frac{\sigma(E_e)}{\sigma(E_i)}n_i + c \commaeq
\end{equation}
where $c$ is an integration constant. Ideally the fraction of the cross sections is one. However due to different simplifications taken in the code to speed up calculation of the cross section for both the electron and ion species this is not the case, but it is nonetheless close to one.\\
The three test case simulations all share the same setup. The electrons and ions are both injected from the left side of the simulation domain and propagate to the right. The ion energy was set to \SI{0.1}{eV} and the electron energy to \SI{9.1e-6}{eV} in order to achieve a similar density in the absence of any interaction. In total a number of \num{8e4} events were simulated.\\
The densities and their corresponding expectations are shown in Figures~\ref{fig:recombination_e_only} and \ref{fig:recombination_t2p_only}. It can be seen that fit quite well with only small statistical uncertainty. From the simulation results the cross sections can be inferred as well. These deviate less than 2\% for both cases from the expected values.\\
The dependence of the electron density on the ion density are shown in Figure~\ref{fig:recombination_ne_nt2p}. Both the ideal slope and the expected slope as defined in equation~\ref{eq:ne_nt2p} can be seen and it is apparent that the data points follow this relation quite well. The expected value of the slope is $\sigma_e/\sigma_{\text{T}_2^+} = 0.973$.
}

\begin{figure}
\centering
\includegraphics[width=1\textwidth]{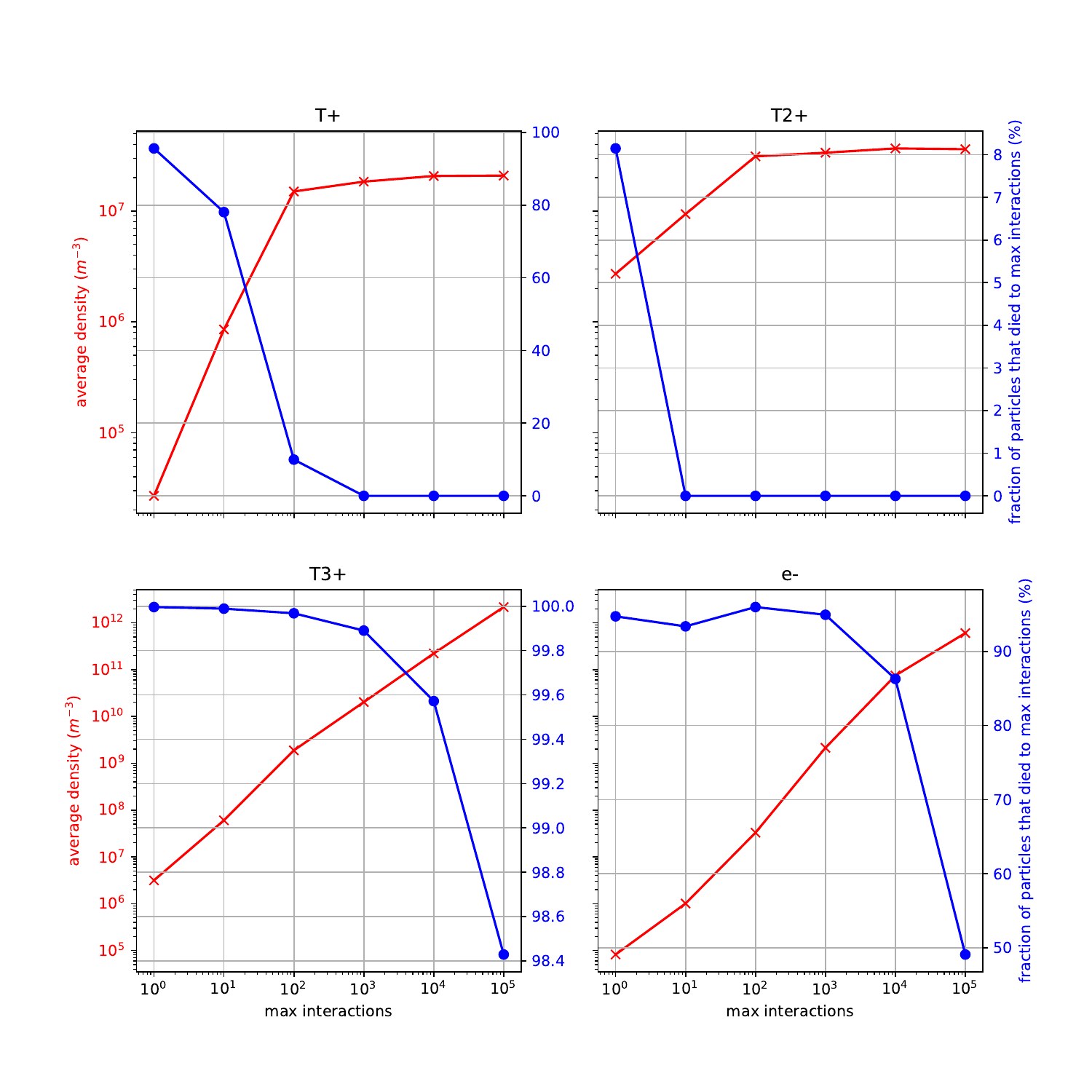}
\caption{Dependence of the densities of different tracked species on the number of maximal interactions. Simulation has been performed using a neutral background specie density of \SI{2.41e24}{\per\cubic\metre}. The densities of the shown species have been averaged over the whole simulation domain.}
\label{fig:maxInteractions_vs_density_2e24}
\end{figure}

\begin{figure}
\centering
\includegraphics[width=1\textwidth]{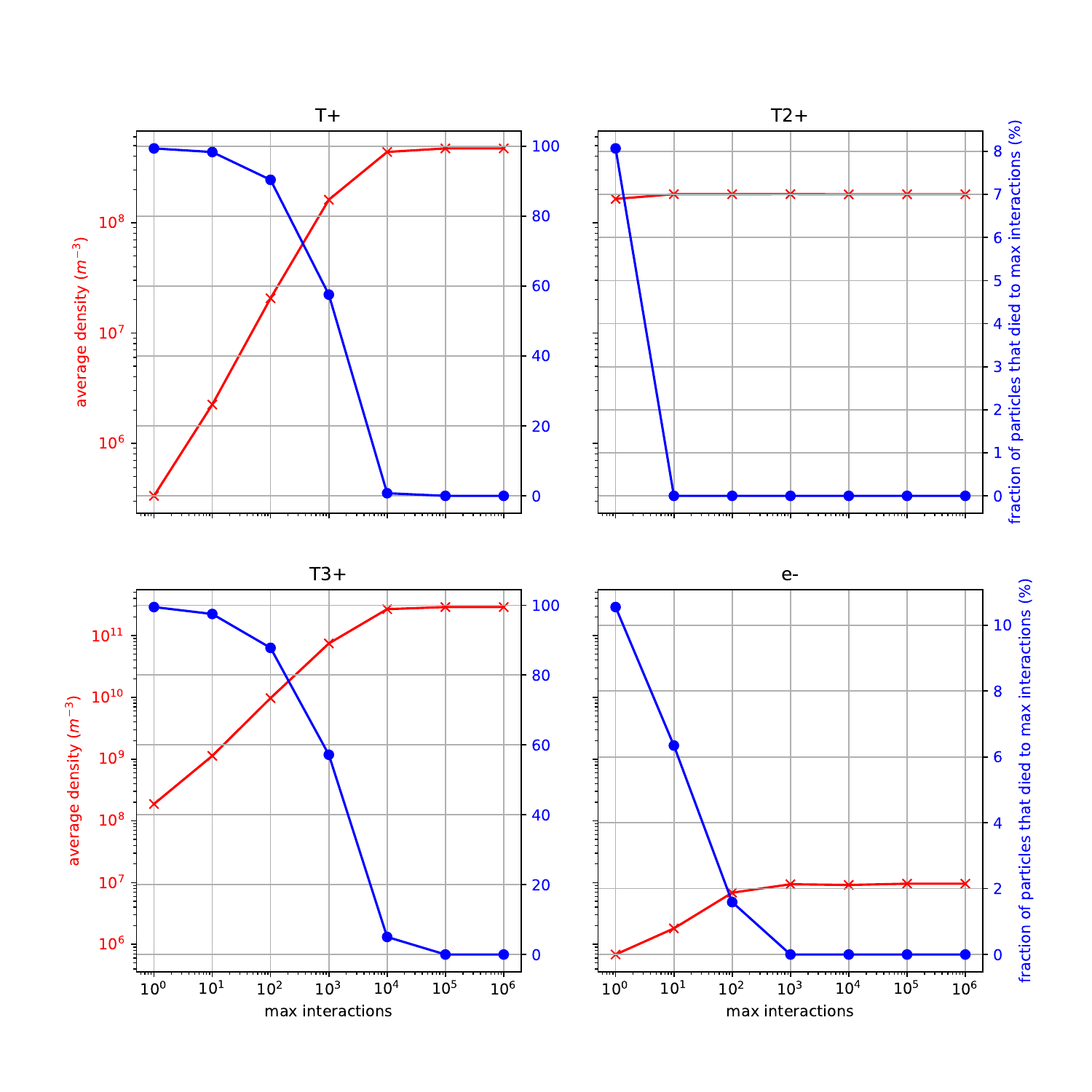}
\caption{Dependence of the densities of different tracked species on the number of maximal interactions. Simulation has been performed using a neutral background specie density of \SI{2.41e22}{\per\cubic\metre}. The densities of the shown species have been averaged over the whole simulation domain.}
\label{fig:maxInteractions_vs_density_2e22}
\end{figure}

\paragraph{Maximum Interaction Count}
\changed{Removing particles form the simulation before they physically leave the domain or are lost due to an interaction introduces an error in the simulation. However, it can be a necessary process in order to make the simulation time be sensible. It is nonetheless necessary to investigate the influence of the maximum number of interactions on the density of the simulated particles.\\
It is obvious that the optimal number of maximum interactions depends on multiple factors, i.e. the electric and magnetic background fields as well as the mean free path of the particle. Since the presence of electromagnetic fields that cause the trapping of a particle is relatively unlikely to occur on their own, the neutral density will have the major impact on the time a particle spends in the simulation. \\
Therefore two sets of simulations with a different neutral density $n_n$ were performed. Apart from the number of maximum interactions and the neutral density, the remaining parameters were not changed. The simulation domain was chosen to have a radius $R = \SI{0.4}{cm}$ and a length of $L = \SI{3}{cm}$. The temperature for the background gas was chosen to be at \SI{300}{\K}. The reason being that in a small simulation domain particles terminate quicker at one of the boundary walls and thus speed up computation time a bit. For the same reason it was chosen to not use a magnetic field. All implemented interactions were enabled during the simulation. The two chosen neutral density values are \SI{2.41e24}{\per\cubic\m} and \SI{2.41e22}{\per\cubic\m}.\\ %mention total simulated particles for each value and how the points are calculated and the density distribution of the neutral gas
Figure~\ref{fig:maxInteractions_vs_density_2e24} shows the density and the fraction of particles that are deleted due to reaching their maximum number of interactions for the \SI{2.41e24}{\per\cubic\m} case. It can be seen that for the T$_2^+$ population a maximum interaction count of about 10 would suffice for all particles to terminate before reaching that limit. This is due to the high clustering cross section with the background gas. Both for T$^+$ and T$_2^+$ the densities converge relatively quickly, i.e. within a limit of \num{1e3} interactions. Some fluctuation is still visible afterwards but that is likely due to more electrons being able to ionize and thus creating more of these species. For both T$_3^+$ and e$^-$ the simulated limit of \num{1e5} maximum interactions is too low for the density to have converged or the fraction of particles that are deleted due to the interaction restriction to have reached zero. This effect is especially pronounced for the T$_3^+$ population due to the high elastic scattering probability at thermal energies. \\
Figure~\ref{fig:maxInteractions_vs_density_2e22} shows the same plot as before but now with the second, lower density. The main differences are that this time all densities converge and that it takes longer for T$^+$ to reach convergence. This is due to the reduced cross section for clustering, which increases the effect of interactions in which the ion is not destroyed. For this case a maximum interaction count of \num{1e5} would be sufficient to capture the correct densities.}

\section{Results}
\label{sec:results}

The simulation tool KARL was designed to determine the charged particle distributions inside of the KATRIN source and to investigate influences of external parameters on this spectrum. Some results of charged particle spectra, particle densities and currents at various positions in the source were already presented and discussed in \citep{Kellerer_2022}, including a comparison of results at different source conditions. 

It was calculated using a factor 10 reduced particle density in the source compared to the simulations in \citep{Kellerer_2022}. The specific input file can be found at $\text{simulation\_config}/\text{parameter}\_\text{katrin\_example.json}$. Source parameters for this scenario include the standard geometry $L=\SI{1657}{\centi\meter}$, $R=\SI{4.5}{\centi\meter}$ and a magnetic field of $B=\SI{2.5}{\tesla}$. The maximum neutral tritium density was chosen to be $n_{\text{T}_2}= \SI{6.3E19}{\per\cubic\meter}$. The most important numerical parameters are the maximum interaction number of $10^6$ and a maximal drift distance of $\SI{1e-3}{\meter}$

\begin{figure}[tbp]
  \centering
  \includegraphics[width=0.8\textwidth]{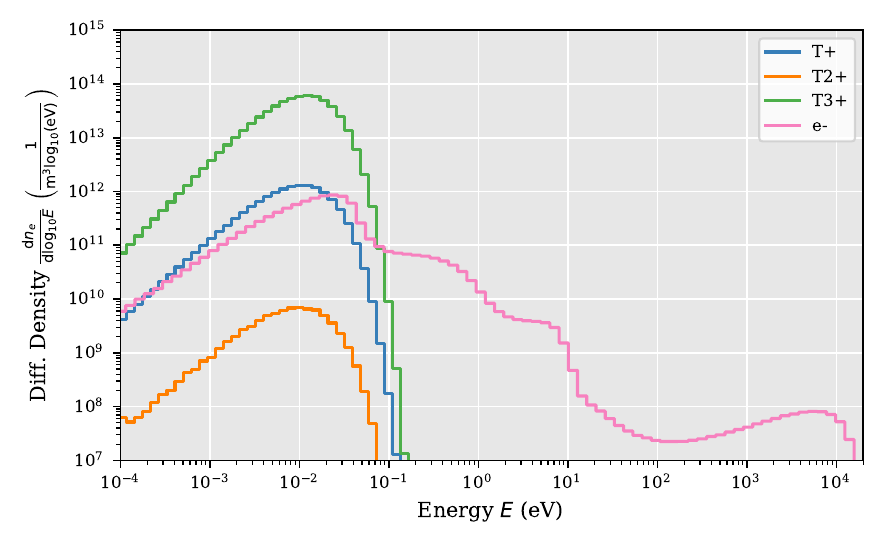}
  \caption{Particle spectra in the center of the source. Simulated using a maximum Tritium density of \SI{6.3E19}{\per\cubic\meter} and a constant magnetic background field of \SI{2.5}{T}.}
  \label{fig:katrin_particle_spectrum}
\end{figure}

It can be seen, that the electron spectrum has a significantly different shape than the spectrum of the ions. It reaches up to values of \SI{18.3}{keV}, while the ion spectra drop off at approximately \SI{0.1}{eV}. This behavior can be attributed to the different creation mechanism and interactions of the particles \citep{Kellerer_2022}.

Electrons are created with energies drawn from the Fermi distribution of $\upbeta$-decay and are then cooled down by ionization and excitation processes. Furthermore electrons can perform elastic scatterings with the neutral gas. The combination of the processes produce the shown spectrum, with the characteristic Fermi distribution at high energies ($> \SI{100}{eV}$), the thermal like distribution at low energies ($< \SI{0.1}{eV}$) and the intermediate distribution including the ionization cutoff at approximately \SI{10}{eV}. 

Ions on the other hand are created with thermal energies of the neutral gas and then perform mostly elastic scatterings, which preserves the Maxwell Boltzmann distribution. The thermal distributions of ions and electrons differ due to different cross sections at lower energies \citep{Kellerer2022_1000143868}. The relative height of the differential density for the different ion species is a result of clustering processes, which will result in the end in T$_3^+$ ions. Clustering is more efficient for T$_2^+$ ions than for T$^+$ ions, which results in a higher density for T$^+$ ions than for T$_2^+$ ions. 

\changed{The spectrum shows that electrons and ions from the beta-decay undergo a number of processes. Otherwise the existence of different types of tritium ions could not be explained. This makes the use of sophisticated Monte Carlo necessary.}

\subsection{Comparison to other Monte Carlo tools}

\begin{figure}[tbp]
  \centering
  \includegraphics[width=0.8\textwidth]{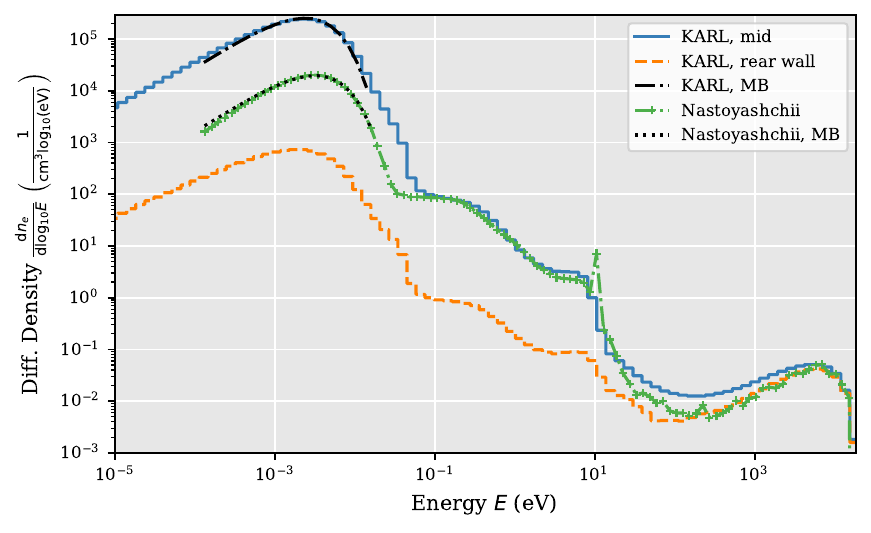}
  \caption{Particle spectra from the KARL code at different positions in the source in comparison with results obtained by \citep{Nastoyashchii:2005aa}. As Nastoyashchii et al. only calculate spectra by taking into account the lifetime within the source there is no possibility to derive spectra at a defined position. The spike at around  \SI{10}{eV} is due to an incorrect representation of excitation processes. }
  \label{fig:Nastoyashchii_comp}
\end{figure}

\changed{To our knowledge KARL is the only freely available Monte Carlo simulation for tritium based experiments. There are however other attempts to model the processes inside the KATRIN experiment. The earliest example is the paper by Nastoyashchii et al. \citep{Nastoyashchii:2005aa}. They follow the idea of a Monte Carlo code where events are generated from the decay of the tritium gas within the source. Most of the elastic and inelastic scattering processes which are implemented in the KARL code are also present in the code by Nastoyashchii et al. with recombination being the remarkable exemption. The most striking difference however is the lack of spatial resolution: The only result of the code is the lifetime of particles in the source (and the spectrum derived from that). Due to the extreme inhomogeneity of the tritium density the results produced by this approach are in many cases not applicable. The results of KARL are compared to this approach in Fig. \ref{fig:Nastoyashchii_comp}.}

\changed{Previously further simulations for the KATRIN experiment have been published \cite{TUMMC}. This code also includes many of the processes included in KARL but lacks the tracing of secondary particles completely and focusses solely on the impact of tritium interactions on the primary particles.}

\subsection{\cchanged{Further results for the KATRIN experiment}}

\cchanged{KARL has been developed with specific applications in the context of the KATRIN experiment in mind. From the physics point of view the following questions were mainly of interest: 1) What is the particle distributions at any point within the source? This is an input quantity for further plasma simulations that require spatially resolved data. 2) What is the current through the rear wall? This is one of the few plasma physics observables of the experiment. 3) What is the current through the radial wall? Observations of the current suggest missing electrons and scattering processes close to the wall.}

\cchanged{Inspired by these questions new features have been developed for KARL, primarily the ability to calculate spectra and densities at any (predefined) point within the source. Due to the extreme inhomogeneity of the tritium density this is a very important feature which may not play such a decisive role in other Monte Carlo tools like GEANT which are primarily used for rather homogenous target materials. Results for the physical questions at hand will be presented below.}
 
%The absolute current to the tube is also influenced by the density profile of electrons and ions. There is no analytic expression of the radial currents. It can be seen that the shape of the current towards the beam tube is similar to the shape of the neutral particle density. This is caused by the increased elastic scattering probability in the center due to the increased neutral gas density. The T$^+$ current is below the T$_3^+$ current, which is caused by the density difference and the different Larmor radius of the two ion species. The current of T$^+$ is in the same order of magnitude than the current of the electrons, even though the masses of both species are significantly different. This behavior is caused by the large difference in density. There is a contribution of T$_2^{+}$ to the radial current. This means that there must be a non-negligible amount of charge transfer reactions, or that enough T$_2^{+}$ is created close to the beam tube walls. The shape of the T$_2^+$ current does not coincide perfectly with the tritium density. Thus, it must be assumed in first order that both effects play a role.

\subsubsection{\cchanged{Position dependent spectra}}

\begin{figure}[tbp]
  \centering
  \includegraphics[width=0.8\textwidth]{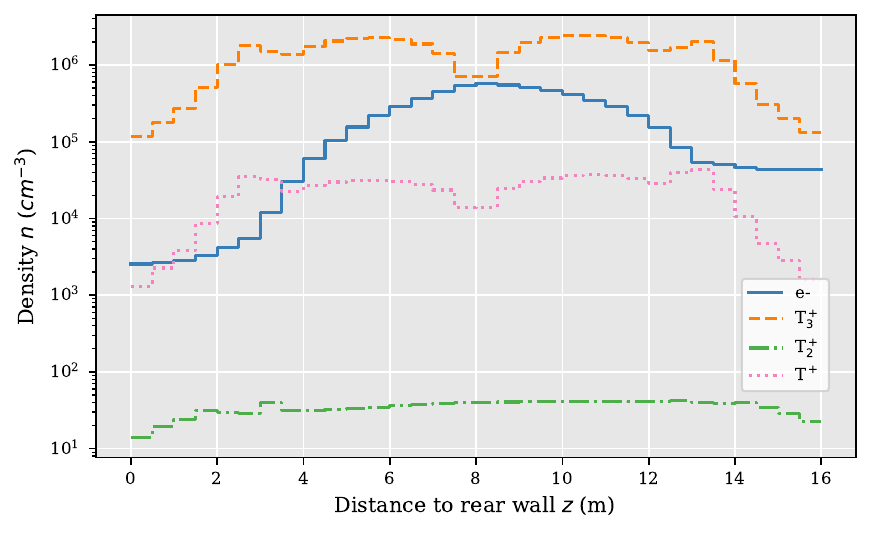}
  \caption{Densities for different species simulated by the KARL code. The densities are measured at predefined interfaces at fixed $z$. The plateaus mark the longitudinal bin width,   where the total current flows through. The simulation is using the standard KATRIN scenario as discussed before. }
  \label{fig:densities_all_vs_z}
\end{figure}

\changed{The composition of different species with respect to the position along the $z$-axis is one example of the outputs of the KARL code. Fig. \ref{fig:densities_all_vs_z} shows the graph where one can see that the electron density resembles the tritium gas density, while T$_2^+$ has a flat distribution. T$_3^+$ as a product of clustering has its density maximum also farther out. In general the higher clusters have a much higher density than the one- and two-atomic ions.}

\subsubsection{\cchanged{Position dependent currents}}

\cchanged{Charged particle currents within the KATRIN tritium source are relevant as a diagnostic parameter since the current at the end of the WGTS can be measured. Additionally ion currents are an input quantity for subsequent plasma simulations. }

\begin{figure}[tbp]
  \centering
  \includegraphics[width=0.8\textwidth]{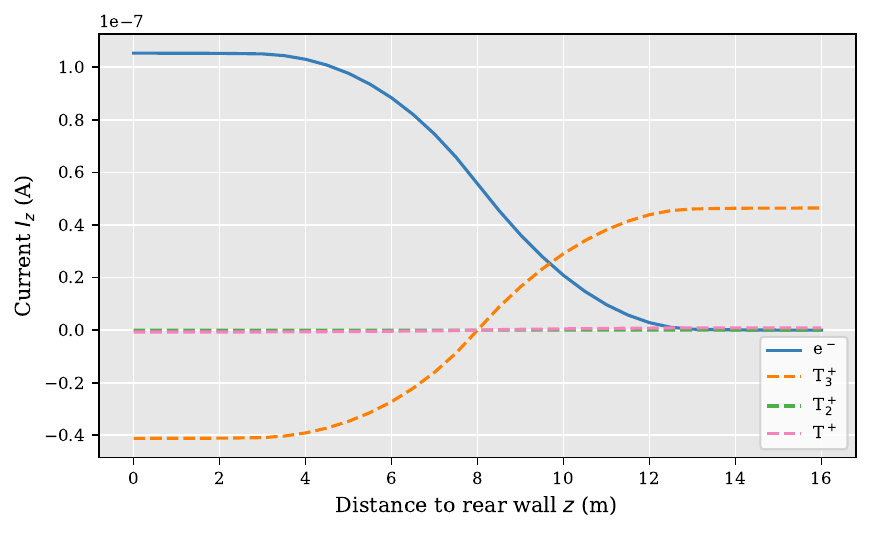}
  \caption{Longitudinal particle currents. The sign of the current corresponds to  the direction of the particle flow multiplied with the sign of the particle charge. Tritium
  is injected at 8 m distance to the rear wall.}
  \label{fig:Iz_vs_z}
\end{figure}

\changed{
The longitudinal particle current was determined through 33 virtual planes in longitudinal direction. The planes were aligned in such a way that the first and last longitudinal plane coincide with the rear wall and}\cchanged{ the transition to the detector.}\changed{ The corresponding longitudinal current profile can be found in Fig. \ref{fig:Iz_vs_z}.
}

\changed{
 The shape of the electron current is as expected. Electrons are reflected at the DPS and the corresponding current is therefore zero. Electrons are most likely to be terminated at the rear wall. Thus, the electron current is the highest here. The electron current decreases continuously from the rear wall towards the DPS. Hence, the electrons can move from the upstream side towards the downstream side and are not hindered from passing by the neutral gas flow.
}

\changed{
The ions are created in the center and stream to both sides. There is no driving force of the ions towards the other side of the inlet. Thus, the ion current at the inlet (\SI{8}{m}) is zero. The current heading towards the DPS and towards the rear wall is almost identical. The differences can be explained by the accessory tritium gas between the first pump and the rear wall.
}

\begin{figure}[tbp]
  \centering
  \includegraphics[width=0.8\textwidth]{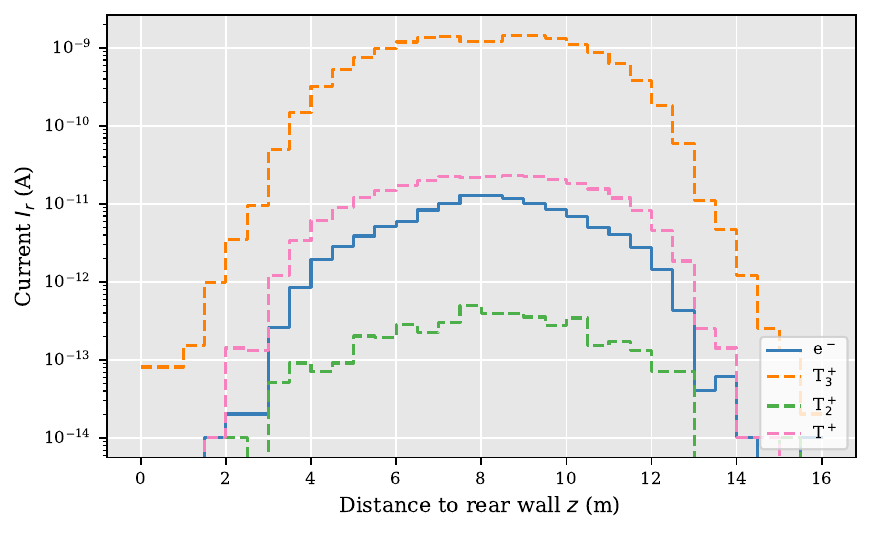}
  \caption{Source tube currents. Absolute radial particle current towards the
  beam tube in longitudinal direction. The plateaus mark the longitudinal bin width,   where the total current flows through. Standard KATRIN parameters are used.}
  \label{fig:tube_current_vs_z}
\end{figure}

\changed{
The sum of the ion current leaving the source at the DPS and at the rear wall is below the electron current leaving at the rear wall. This can only be explained through an ion current, which is directed towards the beam tube. The  corresponding current was detected through a radial virtual plane located at the beam tube walls. The corresponding current can be found in Fig. \ref{fig:tube_current_vs_z}. This current is caused by the change of position of the guiding center after elastic scattering. The difference in position is dependent on the Larmor radius of the particle, which in turn scales with the square root of the mass. Additionally, the cross section of ion scattering is much higher than for electron scattering. Therefore, more collisions occur, which shift the guiding center. Thus, the radial ion current is expected to be significantly higher than the radial electron current.
}

\changed{
The absolute current to the tube is also influenced by the density profile of electrons and ions. There is no analytic expression of the radial currents. It can be seen that the shape of the current towards the beam tube is similar to the shape of the neutral particle density. This is caused by the increased elastic scattering probability in the center due to the increased neutral gas density. The T$^+$ current is below the T$_3^{+}$ current, which is caused by
the density difference and the different Larmor radius of the two ion species. The current of T$^+$ is in the same order of magnitude than the current of the electrons, even though the masses of both species are significantly different. This behavior is caused by the large difference in density. There is a contribution of T$_2^{+}$ to the radial current. This means that there must be a non-negligible amount of charge transfer reactions, or that enough T$_2^{+}$
is created close to the beam tube walls. The shape of the T$_2^+$ current does not coincide perfectly with the tritium density. Thus, it must be assumed in first order that both effects play a role.
}

\changed{
The magnetic field strength can be adjusted in the experiment. In most of the previous measurement runs, the field value is set to $B = \SI{2.5}{\tesla}$. Nevertheless, the field strength can be lowered for commissioning measurements. Simulations showed that the magnetic field strength has no significant influence on the electron spectrum. Nevertheless, it can have an influence on the radial movement of charges in elastic scattering and therefore on
the particle currents. The movement in radial direction is governed by the movement per collision, which is dependent on the gyro radius. The gyro radius in turn is influenced by the magnetic field. 
}

\begin{figure}[tbp]
  \centering
  \includegraphics[width=0.8\textwidth]{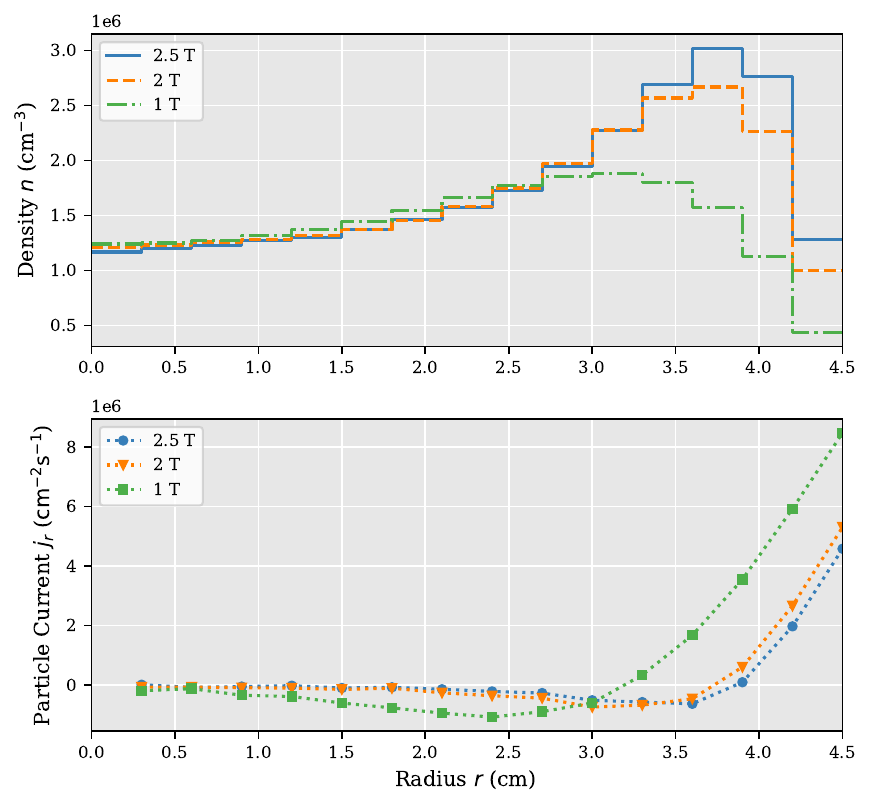}
  \caption{Current in radial direction as a function of radius using the standard KATRIN setup with varying background magnetic fields. }
  \label{fig:comparison_j_n_vs_r}
\end{figure}

\changed{Fig. \ref{fig:comparison_j_n_vs_r} shows the corresponding graph for varying magnetic fields. It is evident that the radial current is larger for smaller magnetic fields. A closer analysis suggests that the current is associated with diffusion. With increasing magnetic field the Larmor radius of particles decreases making it less probable to move perpendicular to the magnetic field line.}

\section{Build instructions}
In order to compile the source code of KARL, download the code on a Unix system and follow the instructions in the README file. The necessary dependencies are listed in the REQUIRED file. A sample configuration file (test\_parameter.json) can be found in the KARL folder simulation\_config.

\section*{Acknowledgments}

% Nemo
The authors acknowledge support by the state of Baden-Württemberg through bwHPC and the German Research Foundation (DFG) through grant no INST 39/963-1 FUGG (bwForCluster NEMO).

J.K. acknowledges the support of the Ministry for Education and Research BMBF (5A17PDA, 05A17PM3, 05A20VK3).

F.S. acknowledges the support of the Deutsche Forschungsgemeinschaft through grant SP 1124/9.

The authors would like to thank Christian Reiling for his work on earlier versions of the KARL code.

\cchanged{We are grateful for the critical comments by the anynomous referees which helped improving the article greatly.}
% ?
%MPI 1.1 support was provided by the Open MPI implementation under the BSD-new license. Design and implementation of the MPI standard by OpenMPI is described in \cite{Gabriel_2004}.

% von dir Felix?
% This work is based upon research supported by the National Research Foundation and Department of Science and Technology. Any opinion, findings and conclusions or recommendations expressed in this material are those of the authors and therefore the NRF and DST do not accept any liability in regard thereto.

%\section{References}
\bibliographystyle{elsarticle-num}
\bibliography{paper}
\end{document}